\documentclass[11pt]{article}

\usepackage{float}
\usepackage{cite}
\usepackage{enumitem}
\usepackage[version=4]{mhchem}
\usepackage{graphicx}
\usepackage[font=tiny]{caption}
\usepackage{authblk}
\usepackage[colorlinks=true, citecolor=blue, linkcolor=blue, urlcolor=blue]{hyperref}
\usepackage[numbers]{natbib}
\title{Advanced X-rays techniques for research-oriented high-resolution imaging of articular cartilage: a scoping review}

\date{}

\author[1]{Simone Fantoni}
\author[2,3]{Luca Brombal}
\author[4]{Paolo Cardarelli}
\author[1]{Fabio Baruffaldi}

\affil[1]{Medical Technology Laboratory, IRCCS Istituto Ortopedico Rizzoli, via di Barbiano 1/10, Bologna, 40136, Italy}
\affil[2]{Department of Physics, University of Trieste, via A. Valerio 2, Trieste, 34127, Italy}
\affil[3]{INFN, Division of Trieste, via A. Valerio 2, Trieste, 34127, Italy}
\affil[4]{INFN, Division of Ferrara, via G. Saragat 1, Ferrara, 44122, Italy}

\begin{document}

\maketitle




\begin{abstract}
Articular cartilage is a musculoskeletal soft tissue renowned for its unique mechanical properties. Understanding both its hierarchical structure and the interplay between its constituents could shed light on the mechanical competence of the tissue. Therefore, rheologic approaches based on high-resolution non-destructive imaging techniques are desired. In this context, X-ray imaging could ideally accomplish this task.

Nevertheless, the nature of articular cartilage translates into poor contrast using conventional absorption modality. To overcome this limitation, several approaches can be embraced. X-ray visibility of articular cartilage can be increased with the use of radiopaque contrast agents. Therefore, further discrimination of structures could be provided by spectral techniques, pivoting on either multi-energy acquisitions or photon-counting technology.

Alternatively, phase-contrast techniques unveil details typically undetected with conventional approaches. Phase-contrast imaging, based on the intrinsic decrement in the refractive index of the tissue, can be achieved with different configurations and implementations, including distinct X-ray sources and optical elements.
Additionally, some phase-contrast techniques retrieve the small-angle scattering-based dark-field signal, relatable to sub-pixel structures. This scoping review aims to catalogue the application of these advanced X-ray techniques to articular cartilage imaging, following PRISMA guidelines. It discusses their advantages, limitations, and includes an overview of rheologic applications to articular cartilage.
\end{abstract}




\subsubsection*{Abbreviations}
\begin{description}[style=nextline,leftmargin=2cm,labelindent=0.5cm]
\item[ABI] 		Analyzer-Based Imaging
\item[AC] 		Articular Cartilage
\item[CA] 		Contrast Agent
\item[CT] 		Computed Tomography
\item[CECT] 		Contrast-Enhanced Computed Tomography
\item[DE] 		Dual-Energy
\item[DF] 		Dark-Field
\item[DVC] 		Digital Volume Correlation
\item[ECM] 		ExtraCellular Matrix
\item[EI] 		Edge-Illumination
\item[GI] 		Gratings Interferometry
\item[microCT] 	micro-Computed Tomography
\item[MIR] 		Multiple-Image Radiography
\item[MRI] 		Magnetic Resonance Imaging
\item[NP] 		Nanoparticle
\item[PBPC] 		Propagation-Based Phase-Contrast
\item[PC] 		Phase-Contrast
\item[PCD] 		Photon-Counting Detector
\item[PG] 		Proteoglycan
\item[PMA]		Phosphomolybdic Acid
\item[PTA]		Phosphotungstic Acid
\item[SR] 		Synchrotron Radiation
\item[USAXS] 		UltraSmall-Angle X-ray Scattering
\end{description}

\section{Introduction}

Many biomedical imaging techniques allow for non-invasive evaluation of heterogeneous tissues. The osteoarticular tissue represents an excellent case of study, as it features different compositions and structures. Besides its crucial role in clinical routine, imaging of osteoarticular tissues has been extensively investigated in the research frame. In the latter context, the retrieved information is associated with the functional properties of osteochondral tissues. Their degradation, related to pathologies affecting the articulating joints, can be easily depicted with computed tomography (CT) and magnetic resonance imaging (MRI). The role of CT and MRI has been extensively discussed in the context of clinical diagnostics. However, such discussions are limited to conventional techniques, as diagnostic imaging performed on patients prioritizes minimizing the acquisition times, the possible hazards and the costs. Conversely, biomedical research places more emphasis on different aspects, such as spatial resolution and quantitative accuracy. In this frame, the high-resolution counterpart of CT, the microcomputed tomography (microCT), is preferred.

MicroCT can provide detailed information on microstructures of biological tissues \cite{Mizutani2012X}, even down to the cellular level, with results comparable to histology \cite{Zehbe2010Going}. On the other hand, nearly no contrast of soft tissues such as articular cartilage (AC) is achieved, due to their associated low atomic numbers Z. Consequently, limited information is extracted with conventional X-ray techniques, reducing their potential in assessing AC status. The need for accurate examination of AC and soft tissues, crucial for early disease detection and biomedical analysis, has driven the exploration of alternative X-ray imaging techniques, based on the implementation of non-conventional X-ray sources, detectors, contrast agents (CAs), or their combination.

The aim of the present review is to report on developments achieved in the frame of high-resolution X-ray imaging of AC in recent years, drawing attention to innovative techniques, implemented with different imaging systems.

\subsection{Articular cartilage}

\begin{figure}[!h]
\centering\includegraphics[width=4in]{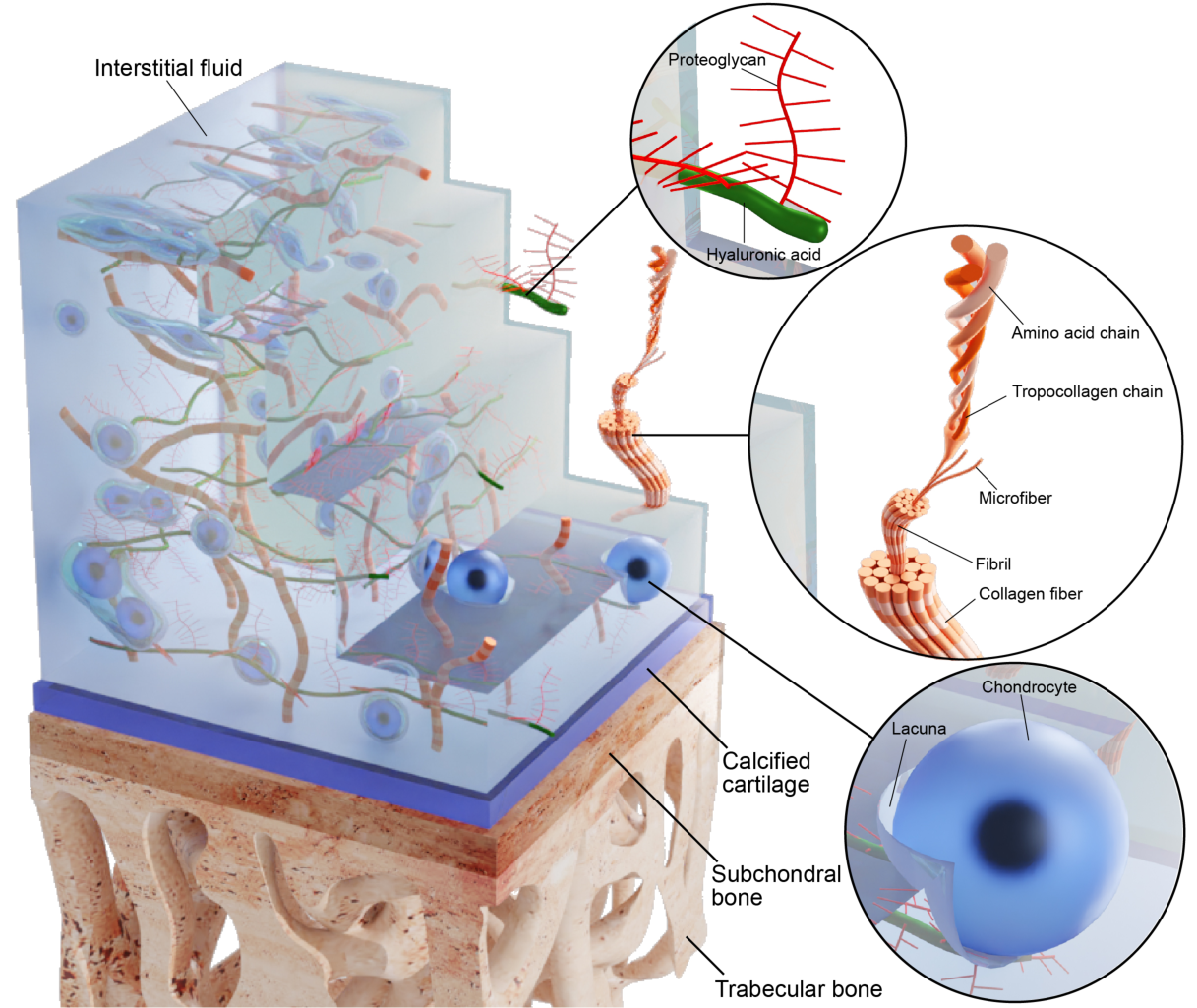}
\caption{Scheme of AC}
\label{fig_intro}
\end{figure}

Articular cartilage is a connective tissue covering joint surfaces, mainly composed of collagen fibrils (predominantly type II), proteoglycans (PGs), and interstitial fluid. Originating from the deeper zone of the tissue, the collagen fibrils arrange into bundles surging aligned to AC tissue surface, acquiring the characteristic arcade-like structure. The PGs (i.e., the non-fibrillar component of the ECM) are complex proteins bound with multiple glycosaminoglycans chains, and mainly originate from the pericellular environment of chondrocytes. Their concentration follows a peculiar depth-wise gradient, increasing towards the deep zone of AC. The chondrocytes distribute inside the tissue in specific sites (i.e. lacunae) and synthetize the main components of ECM. The structure of AC is schematized in Figure \ref{fig_intro}.

The residual negative charge density displayed by glycosaminoglycan chains confers unique properties to the tissue. For instance, the entwined arrangement of PGs in between the collagen bundles generates a spatial outstretch of the ECM, owing to the mutual repulsion of negative charge densities. This effect gives rise to the mechanical pre-tensioning of the AC. Moreover, the charge imbalance draws cations dissolved in the interstitial fluid. The latter results in Donnan osmotic pressure and induces a significant hydration and swelling of the tissue. This phenomenon, together with tissue pre-tensioning and the fibril solid component of ECM, accounts for the characteristic instantaneous response of AC to any rapid mechanical solicitation. The porosity associated with the packed ECM organization regulates the transient reconfiguration of interstitial fluid following any mechanical stimulus, until equilibrium is reached \cite{Mow1992Cartilage,Ateshian2009role}. In summary, the hierarchical structure and composition of AC translate into near-frictionless and smooth joint motion, provided the integrity of the tissue \cite{Mow1992Cartilage}.

The degeneration of ECM, typically occurring with ageing, after traumas or associated with pathologies (i.e. osteoarthritis), alters both the composition and structure of AC \cite{Setton1999Altered}. More in detail, the fibrillation of collagen bundles, the depletion of PGs, and the augmented content of interstitial fluid in ECM are recognized as hallmarks of AC diseases, along with modifications of underlying subchondral bone as well \cite{Mahjoub2012Why}. The reduced mechanical performance of AC induces the progressive impairment of joint motion, with severe compromission of an individual’s mobility.

In the clinical frame, the early detection of subtle changes in AC could improve the prevention of related musculoskeletal diseases. X-ray imaging methods allow for a detailed depiction of both soft and mineralized tissues at high spatial resolution in restrained acquisition times. For instance, the link between the tissue microstructure and its mechanical response has been investigated to relate the different degradation stages of AC with the tissue functionality. In-situ testing enables the study of geometrical, morphological, and densitometric properties of biological tissues in conjunction with their mechanical properties. In this context, digital volume correlation (DVC) can track down these subtle interrelationships by measuring the full three-dimensional displacement and strain maps under prescribed loading conditions. Nonetheless, the accuracy and precision of the DVC in measuring displacements strongly depend on the quality of the input images (i.e., signal-to-noise ratio and spatial resolution) \cite{DallAra2017Precision}.

As it will be discussed in the review, different mechanisms of image formation yield quantitative information of AC properties, including tissue morphology and composition. The multiple nature of signals (namely absorption, refraction, and scattering of radiation) that can be extracted with advanced X-ray techniques provides meaningful results, as supported by independent methods in the selected literature, and lay the foundations for their use in AC imaging, as introduced hereafter.

\section{Advanced X-ray imaging techniques}

In general X-ray imaging techniques can give access to three signals, namely absorption, refraction or scattering phenomena. Absorption-based X-ray methods are easily performed with conventional systems, but result in low contrast of soft tissues due to their low electron density. The use of high-Z CAs significantly enhances the visibility of inner structures of AC. In this context, to further enhance the discrimination of tissues sharing comparable radiopacity, spectral imaging techniques make use of the specific energy dependence of CA attenuation coefficients. Owing to their spectroscopic capabilities, photon-counting detectors (PCDs) are a crucial implementation for spectral imaging \cite{Mourad2024Chances}. Refraction-based methods exploit the optical phenomena originating from the shift in the phase of X-ray waves accumulated while traversing the sample. This, in combination with specific geometrical configurations or the insertion of optical elements, allows for the detection of a signal from soft tissue features, generally undetectable with absorption-based approaches. Additionally, valuable information on tissues’ substructures can be retrieved by assessing the scattering of X-rays in the sample, which can be modelled as a multiple-refraction process at a scale smaller than the system’s spatial resolution.
Advanced X-ray imaging techniques, allowing for the collection of the different signals previously introduced, can be implemented with laboratory or synchrotron systems. The first category, including conventional microCT scanners, is widely spread thanks to the high availability of commercial systems. Nevertheless, laboratory scanners are mainly devoted to absorption-based modalities, owing to low spatial coherence and limited fluxes. Existing literature reports only a few experiences exploiting microfocus X-ray tubes for phase-contrast (PC) imaging of AC \cite{Clark2020Propagation}. Synchrotron radiation (SR)-based systems, featuring high fluxes, monochromaticity and high spatial coherence, find applications in retrieving refraction and scattering signals. On the other hand, the limited accessibility to synchrotron facilities hinders their translation to routine practice.

\subsection{Absorption-based X-ray imaging}
\subsubsection{Contrast-enhanced computed tomography}

\begin{figure}[!h]
\centering\includegraphics[width=4in]{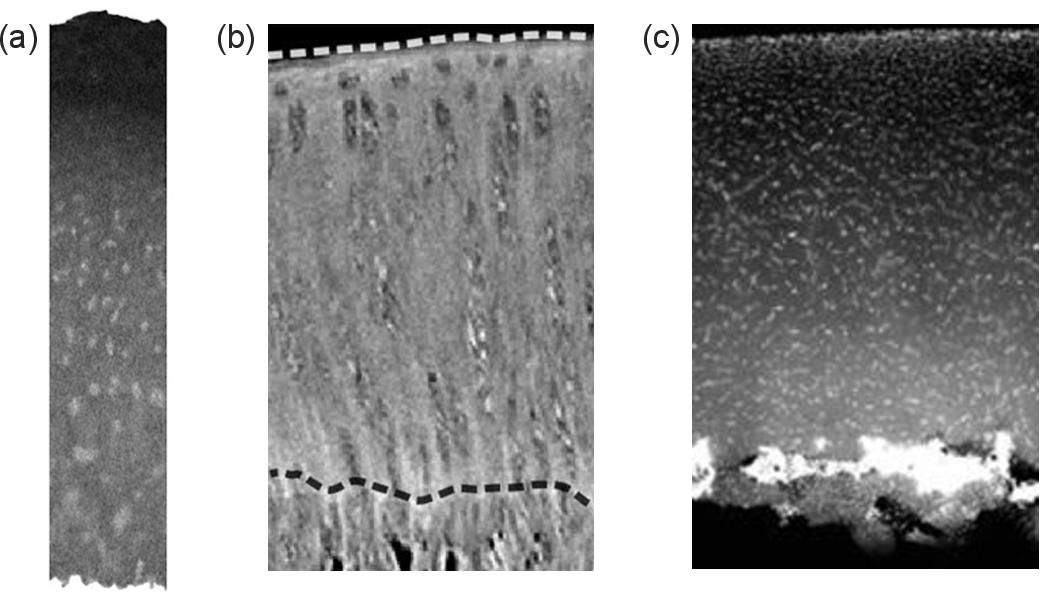}
\caption{(a) Section from contrast-enhanced microCT of human AC, following the immersion in CA4+ solution (3.2 $\mu$m-pixel size, commercial microCT scanner). Along with the characteristic gradient of radiopacity increasing towards the deep layer of AC, spots with augmented signal are recognizable as chondrocytes' lacunae. Image adapted from the work of Karhula et al. \cite{Karhula2017Micro}, published under Creative Commons license (CC BY 4.0). (b) Section from contrast-enhanced microCT of rabbit AC, following the immersion in solution of hexaammineruthenium(III) chloride and cacodylic acid (0.65 $\mu$m-pixel size, commercial microCT scanner). The proposed protocol aimed to investigate the collagen fiber orientation in a osteoarthritic AC model (dashed white line and dashed black line mark tissue surface and tidemark, respectively). Image adapted from the work of Ojanen et al. \cite{Ojanen2023Micro}, published under Creative Commons license (CC BY 4.0). (c) Section from contrast-enhanced microCT of human AC, following the immersion in propidium iodide (2.6 $\mu$m-pixel size, SR-microCT). The direct binding of the selected CA to DNA in chondrocytes allowed the punctual depiction of the cellular pattern, enabling the quantification of chondrocytes' distribution even inside each lacuna. Image adapted from the work of Danalache et al. \cite{Danalache2021Exploration}, published under Creative Commons license (CC BY 4.0).}
\label{fig_intro_CE}
\end{figure}

Contrast-enhanced computed tomography (CECT) is an absorption-based technique that uses exogenous radiopaque contrast agents (CAs). Contrast agents are compounds including high-Z elements, which increase the attenuation coefficient and enhance the contrast of the investigated soft tissue.

First, the selection of CAs depends on the imaging purpose. In-vivo purposes require non-toxic, FDA-approved CAs \cite{Lusic2013X}, whereas the use of heavy metal-based stains \cite{Pauwels2013exploratory} and experimental compounds is limited to ex-vivo and in-vitro studies, due to their high toxicity. Among the available CAs, iodine-based formulations have been extensively used to visualize AC (i.e., ioxaglate, iothalame, iohexol, ioversol), as denoted in Table \ref{tab:table_CE}. For research purposes, CA stains generally employed in histology and transmission electron microscopy have also been translated to X-ray imaging protocols.

\begin{table}[!]
\caption{Summary of research papers including X-ray imaging of AC with CECT protocols, ordered from the most to the least recent. For each paper, the corresponding reference is reported, along with the CA used, its net electric charge $q$, the model for the AC tissue, the size of the specimen $S$, the voltage $V$ of the X-ray beam, the acquisition time $T$ and the voxel size of the reconstructed volume $p$. $\dagger$ SR}
    \centering
    \scriptsize
    \resizebox{\textwidth}{!}{
    \begin{tabular}{cccccccc}
    \hline
        ref. & contrast agent & $q$ (e) & model & $S$ (mm) & $V$ & $T$ (min) & $p$ (µm) \\ \hline
        \cite{Omelchenko2024New} & \ce{Na0.2TiO2} NPs & NA & bovine & NA & 25 keV & NA & 9 \\
        ~ & \ce{H0.3MoO3} NPs & NA & ~ & ~ & 17.5 keV & ~ & ~ \\ \hline
        \cite{Nelson2024Longitudinal} & CA4+ & +4 & equine & Ø 7 & 70 kVp & NA & 36 \\ \hline
        \cite{Jantti2024Cationic} & \ce{Ta2O5} NPs & +1 & equine & Ø 8.5 & 150 kVp & NA & 40 \\ \hline
        \cite{Davis2024Comparison} & PTA & -3 & bovine & Ø 4-10 & 40 kVp & 160 & 3 \\
        ~ & Hf-WD POM & -16 & ~ & ~ & ~ & ~ & ~ \\ \hline
        \cite{Valerio2023Effect} & CA4+ & +4 & rat & NA & 70 kVp & NA & 6 \\ \hline
        \cite{Ojanen2023Micro} & cacodylic acid & NA & rabbit & Ø 2 & 40 kVp & 240 & 0.65 \\
        ~ & \ce{Cl3H18N6Ru} & NA & ~ & ~ & ~ & ~ & ~ \\ \hline
        \cite{Lin2023Intra} & ioxaglate & -1 & rat & NA & 45 kVp & 26 & 16 \\ \hline
        \cite{Jo2023In} & iopamidol & -1 & murine & NA & 55 kVp & 20 & 10.4 \\ \hline
        \cite{Jin2023Effects} & iohexol & 0 & rabbit & Ø 3.5 & 40 kVp & NA & 18.22 \\ \hline
        \cite{Honkanen2023Dual} & ioxaglate & -1 & equine & NA & 90 kVp & 2 & 80 \\
        ~ & bismuth NPs & 0 & ~ & ~ & ~ & ~ & ~ \\ \hline
        \cite{Fleischer2023Early} & ioxaglate & -1 & rat & NA & 55 kVp & NA & 8; 12 \\ \hline
        \cite{Durongbhan2023microCT} & CA4+ & +4 & murine & NA & 70 kVp & 2.9, 3.5 & 10 \\
        ~ & ~ & ~ & chicken & ~ & ~ & ~ & ~ \\ \hline
        \cite{Davis2023Development} & PTA & -3 & guinea pig & NA & 60 kVp & NA & 10.85 \\ \hline
        \cite{Chan2023Contrast} & CA4+ & +4 & murine & NA & 45kVp & NA & 2 \\
        ~ & ~ & ~ & chicken & ~ & ~ & ~ & ~ \\ \hline
        \cite{Bhattarai2023Computed} & Bi-DOTA & +1 & porcine & NA & 70 kVp & NA & 10.06 \\
        ~ & Gd-DOTA & +1 & ~ & ~ & ~ & ~ & ~ \\ \hline
        \cite{McKinney2022Sodium} & ioxaglate & -1 & rat & NA & 45kVp & 27 & 16 \\ \hline
        \cite{Fowkes2022Imaging} & 3,5-diiodo-L- & NA & human & Ø 5 & 70 kVp & 3 & 10 \\
        ~ & tyrosine & ~ & murine & ~ & ~ & ~ & ~ \\ \hline
        \cite{Fantoni2022Cationic} & CA4+ & +4 & bovine & NA & 50 kVp & NA & 11.52 \\ \hline
        \cite{Zhu2021Contrast} & PTA & -3 & bovine & 3-5 & 80kVp & 52/181 & 4.6; 3.4 \\ \hline
        \cite{Ve2021Comparison} & ioxaglate & -1 & primate & NA & 80 kVp & 15 & 90 \\ \hline
        \cite{Nelson2021Quantitative} & CA4+ & +4 & equine & Ø 7 & 70 kVp & NA & 36 \\ \hline
        \cite{Nelson2021Cationic} & CA4+ & +4 & equine & Ø 7 & 70 kVp & NA & 36 \\ \hline
        \cite{Lawson2021Tantalum} & \ce{Ta2O5} NPs & +13/0 & human & NA & 70 kVp & NA & 36 \\ \hline
        \cite{Gao2021Influence} & CA4+ & +4 & rabbit & NA & 70 kVp & NA & 38 \\ \hline
        \cite{Flynn2021Anionic} & iothalamate & -1 & bovine & NA &  80 kVp & 18 & 45 \\ \hline
        \cite{Dunham2021Increased} & PMA & NA & rat & NA & 70 kVp & NA & 15 \\ \hline
        \cite{Dufour2021Repair} & iodixanol & 0 & primate & NA & NA & NA & NA \\ \hline
        \cite{Danalache2021Exploration} & propidium iodide & NA & human & NA & $\dagger$ 12 keV m. & 5 & 2.6 \\ \hline
        \cite{Clark2021High} & PTA & -3 & human & Ø 3 & 50 kVp & NA & 1.8 \\ \hline
        \cite{Boos2021Contrast} & CA4+ & +4 & bovine & Ø 5 & 70 kVp & 27.9 & 4 \\ \hline
        \cite{Blom2021Single} & ioxaglate & -1 & goat & 75 & 70 kVp & NA & 18 \\ \hline
        \cite{Besler2021Quantitative} & \ce{SiO2} & NA & rat & NA & 70 kVp & NA & 10 \\
        ~ & ioxaglate & -1 & ~ & ~ & ~ & ~ & ~ \\ \hline
        \cite{Voert2020Contrast} & ioversol & 0 & mice & Ø 5 & 55 kVp & NA & 57 \\
        ~ & iomeron & 0 & bovine & ~ & ~ & ~ & ~ \\ \hline
        \cite{Sugasawa2020Characterization} & iopimadol & 0 & rat & NA & 47 kVp  & NA & 21 \\ \hline
        \cite{Reece2020Reduced} & ioxaglate & -1 & murine & NA & 45 kVp & NA & 16 \\ \hline
        \cite{Meng2020Diffusion} & iohexol & 0 & human & Ø 16.3 & 84 kVp & 17 & 45 \\ \hline
        \cite{Freedman2020dGEMRIC} & gadopentetate & -2 & human & NA & 70 kVp & 13 & 18 \\
        ~ & Gd(DTPA)Lys2 & +4 & ~ & ~ & ~ & ~ & ~ \\
        ~ & ioxaglate & -1 & ~ & ~ & ~ & ~ & ~ \\
        ~ & CA4+ & +4 & ~ & ~ & ~ & ~ & ~ \\ \hline
	\end{tabular}
	}
	  \label{tab:table_CE}%
  \vspace*{-4pt} 
\end{table}
\begin{table}[H]
    \centering
    \scriptsize
    \resizebox{\textwidth}{!}{
    \begin{tabular}{cccccccc}
    \hline
        ref. & contrast agent & $q$ (e) & model & $S$ (mm) & $V$ & $T$ (min) & $p$ (µm) \\ \hline
        \cite{Cubria2020Evaluation} & ioxaglate & -1 & murine & NA & 70 kVp & NA & 6 \\ \hline
        \cite{Zhang2019Protocol} & CA4+ & +4 & human & NA & NA & NA & 100 \\ \hline
        \cite{Ylitalo2019Quantifying} & PTA & -3 & human & Ø 4 & 80 kVp & 120 & 3.0 \\
        ~ & ~ & ~ & ~ & ~ & 45kVp & ~ & 3.2 \\ \hline
        \cite{Tsai2019Effects} & ioxaglate & -1 & rat & NA & 45 kVp & NA & 14.8 \\ \hline
        \cite{Steppacher2019Ultrasonic} & iopimadol & 0 & bovine & Ø 10 & 70 kVp & NA & 16 \\
        ~ & ~ & ~ & porcine & ~ & ~ & ~ & ~ \\ \hline
        \cite{Nelson2019Evaluation} & CA4+ & +4 & equine & Ø 7 & 70 kVp & NA & 36 \\ \hline
        \cite{Michalak2019Concurrent} & iohexol & 0 & human & NA & 68 kVp & 18 & 61 \\ \hline
        \cite{McKinney2019Therapeutic} & ioxaglate & -1 & rat & NA & 45 kVp & 27 & 16 \\ \hline
        \cite{Kwok2019Knee} & PTA & -3 & murine & NA & 90 kVp & NA & 2.5 \\ \hline
        \cite{Gatenholm2019Spatially} & ioxaglate & -1 & human & NA & 70 kVp & 40 & 36 \\ \hline
        \cite{Dourthe2019Assessment} & CA4+ & +4 & human & NA & 120 kVp & NA & 41 \\ \hline
        \cite{Reece2018Contrast} & ioxaglate & -1 & rat & NA & 45 kVp & 26 & 16 \\ \hline
        \cite{Nickmanesh2018Contrastenhanced} & CA4+ & +4 & human  & NA & 120 kVp & NA & 41 \\ \hline
        \cite{Mirahmadi2018Aging} & ioxaglate & -1 & equine & Ø 6 & 66 kVp & NA & 10 \\
        ~ & iodixanol & 0 & ~ & ~ & ~ & ~ & ~ \\ \hline
        \cite{Stewart2017Synthesis} & CA4+ & +4 & bovine & Ø 7 & 70 kVp & NA & 36 \\
        ~ & ioxaglate & -1 & rat & ~ & ~ & ~ & ~ \\
        ~ & iodixanol & 0 & ~ & ~ & ~ & ~ & ~ \\ \hline
        \cite{Saukko2017Dual} & \ce{Bi2O3} & NA & bovine & Ø 7 & 100 kVp & NA & 15.98 \\
        ~ & ioxaglate & -1 & ~ & ~ & ~ & ~ & ~ \\ \hline
        \cite{Raines2017Efficacy} & meglumine diatrizoate & -1 & rat & NA & 45 kVp & NA & 16 \\ \hline
        \cite{Nieminen20173D} & PTA & -3 & human & Ø 4 & 80 kVp & 100 & 3 \\ \hline
        \cite{Mashiatulla2017Murine} & CA4+ & +4 & murine & NA & 45 kVp & NA & 3 \\
        ~ & ioxaglate & -1 & ~ & ~ & ~ & ~ & ~ \\ \hline
        \cite{Kun2017Contrast} & uranyl acetate & NA & human & NA & 70 kVp & NA & 13; 2.94 \\
        ~ & lanthanide solution & NA & rat & ~ & ~ & ~ & ~ \\
        ~ & IKI & NA & ~ & ~ & ~ & ~ & ~ \\
        ~ & PTA & -3 & ~ & ~ & ~ & ~ & ~ \\
        ~ & iopamidol & 0 & ~ & ~ & ~ & ~ & ~ \\ \hline
        \cite{Karhula2017Effects} & PTA & -3 & bovine & Ø 4.8 & 40 kVp & 8 & 17.4 \\
        ~ & ioxaglate & -1 & ~ & ~ & ~ & ~ & ~ \\ \hline
        \cite{Karhula2017Micro} & CA4+ & +4 & human & Ø 2 & 45 kVp & 122 & 3.2 \\ \hline
        \cite{Visser2017Groove} & ioxaglate & -1 & rat & NA & 90 kVp & 3 & 21 \\ \hline
        \cite{Willett2016Quantitative} & ioxaglate & -1 & rat & NA & 45 kVp & 24.7 & 16 \\ \hline
        \cite{Tiel2016Quantitative} & ioxaglate & -1 & human & NA & 95 kVp & 30-90 & 35 \\ \hline
        \cite{Stok2016Three} & \ce{SiO2} & NA & rabbit & NA & 70 kVp & NA & 18 \\
        ~ & ioxaglate & -1 & ~ & ~ & ~ & ~ & ~ \\ \hline
        \cite{Mittelstaedt2016Topographical} & ioxaglate & -1 & canine & NA & 40 kVp & NA & 13.4 \\ \hline
        \cite{Lakin2016Contrast} & CA4+ & +4 & murine & NA & 70 kVp & NA & 2; 6 \\ \hline
        \cite{Honkanen2016Cationic} & CA2+ & +2 & bovine & Ø 6 & 100 kVp & NA & 25 \\ \hline
        \cite{Wang2015Assessment} & ioxaglate & -1 & rabbit & NA & 45 kVp & NA & 20 \\ \hline
        \cite{Nieminen2015Determining} & PTA & -3 & equine & Ø 6 & 80 kVp & 32 & 8.7 \\ \hline
        ~ & PMA & NA & human & Ø 4.6 & 80 kVp & 135 & 3.2 \\ \hline
        \cite{Li2015Observation} & meglumine diatrizoate & -1 & human & Ø 12 & 50 kVp & 11 & 18 \\ \hline
        \cite{Lakin2015Contrast} & CA4+ & +4 & human & NA & 70 kVp & NA & 36 \\
        ~ & ioxaglate & -1 & ~ & ~ & ~ & ~ & ~ \\ \hline
        \cite{Fan2015Correlations} & meglumine diatrizoate & -1 & bovine & 4 & 50 kVp & NA & 18 \\ \hline
        \cite{Bagi2015Correlation} & ioxaglate & -1 & rat & NA & 55 kVp & 42 & 10 \\ \hline
        \cite{Bagi2015Effect} & ioxaglate & -1 & rat & NA & 55 kVp & 42 & 10 \\ \hline
        \cite{Siebelt2014Inhibited} & ioxaglate & -1 & rat & NA & 65 kVp & NA & 18 \\ \hline
        \cite{Siebelt2014FK506} & ioxaglate & -1 & rat & NA & 65 kVp & NA & 18 \\ \hline
        \cite{Renders2014Contrast} & ioxaglate & -1 & human & NA & 70 kVp & 36 & 18 \\
        ~ & ~ & ~ & murine & ~ & ~ & ~ & ~ \\ \hline
	\end{tabular}
	}
\end{table}
\begin{table}[!h]
    \centering
    \scriptsize
    \resizebox{\textwidth}{!}{
    \begin{tabular}{cccccccc}
    \hline
        ref. & contrast agent & $q$ (e) & model & $S$ (mm) & $V$ & $T$ (min) & $p$ (µm) \\ \hline
        \cite{Kerckhofs2014Contrast} & ioxaglate & -1 & murine & NA & 60 kVp & 20 & 2 \\ \hline
        \cite{Fu2014Impaired} & ioxaglate & -1 & rat & NA & 60 kVp & NA & 20 \\ \hline
        \cite{Freedman2014Tantalum} & \ce{Ta2O5} NPs  & NA & human & NA & 50 kVp & NA & 6; 36; 100 \\
        ~ & ioxaglate & -1 & rat & ~ & ~ & ~ & ~ \\ \hline
        \cite{Entezari2014Effect} & iothalamate & -1 & bovine & NA & 58 kVp & NA & 70; 100 \\ \hline
        \cite{Das2014Rapid} & PTA & -3 & murine & NA & 50 kVp & NA & 5 \\ \hline
        \cite{Thote2013Localized} & ioxaglate & -1 & rat & NA & 45 kVp & NA & 16 \\ \hline
        \cite{Stewart2013Contrast} & CA4+ & +4 & bovine & Ø 7 & 70 kVp & NA & 36 \\
        ~ & ioxaglate & -1 & rabbit & ~ & 50 kVp & ~ & 100 \\ \hline
        \cite{Lakin2013Cationic} & CA4+ & +4 & bovine & Ø 7 & 70 kVp & NA & 36 \\ \hline
        \cite{Wang2012Evaluation} & meglumine diatrizoate & -1 & rat & NA & 70 kVp & NA & 18 \\ \hline
        \cite{Van2012CT} & ioxaglate & -1 & human & NA & 55 kVp & 360-600 & 35 \\ \hline
        \cite{Kotwal2012Initial} & ioxaglate & -1 & murine & NA & 45 kVp & NA & 6 \\ \hline
        \cite{Kerckhofs2012Contrast} & ioxaglate & -1 & murine & NA & 60 kVp & 20 & 2 \\ \hline
        \cite{Siebelt2011Quantifying} & ioxaglate & -1 & rat & NA & 55 kVp & NA & 35 \\ \hline
        \cite{Siebelt2011Clinically} & ioxaglate & -1 & human & NA & 55 kVp & 360-600 & 35 \\ \hline
        \cite{Bansal2011Contrast} & CA4+ & +4 & bovine & Ø 7 & 70 kVp & NA & 36 \\
        ~ & ioxaglate & -1 & ~ & ~ & ~ & ~ & ~ \\
        ~ & gadopentetate & -2 & ~ & ~ & ~ & ~ & ~ \\ \hline
        \cite{Bansal2011Cationic} & CA4+ & +4 & bovine & Ø 7 & NA & NA & 70; 36 \\
        ~ & CA1+ & +1 & rabbit & ~ & ~ & ~ & ~ \\
        ~ & iothalamate & -1 & ~ & ~ & ~ & ~ & ~ \\ \hline
        \cite{Xie2010Nondestructive} & ioxaglate & -1 & rat & NA & 45 kVp & NA & 12; 16 \\ \hline
        \cite{Bansal2010Contrast} & iothalamate & -1 & bovine & Ø 7 & NA & NA & 70; 100 \\ \hline
        \cite{Xie2009Quantitative} & ioxaglate & -1 & rat & NA & 45 kVp & NA & 12 \\ \hline
        \cite{Taylor2009Comparison} & gadopentetate & -2 & human & NA & 60 kVp & NA & 41 \\ \hline
        \cite{Joshi2009Effect} & CA1+ & +1 & rabbit & NA & 45 kVp & NA & 30 \\
        ~ & CA2+ & +2 & ~ & ~ & ~ & ~ & ~ \\
        ~ & CA4+ & +4 & ~ & ~ & ~ & ~ & ~ \\
        ~ & iothalamate & -1 & ~ & ~ & ~ & ~ & ~ \\
        ~ & ioxaglate & -1 & ~ & ~ & ~ & ~ & ~ \\ \hline
        \cite{Piscaer2008In} & ioxaglate & -1 & rat & NA & 55 kVp & 15 & 35 \\ \hline
        \cite{Palmer2006Analysis} & ioxaglate & -1 & bovine & Ø 4 & 45 kVp & NA & 21 \\
        ~ & ~ & ~ & rabbit & ~ & ~ & ~ & ~ \\ \hline
    \end{tabular}
    }
\end{table}

Second, the effectiveness in enhancing AC X-ray visibility relies upon the properties of the CA molecule, namely its size, the modality of interaction, and more importantly, its net electric charge. These properties account for the steric hindrance, the reversibility of the contrast enhancement process and the affinity of CA molecule to the AC, respectively. The molecule size should be tailored to the tissue porosity, to facilitate the permeation in the tissue \cite{Majda2017New}.

The net electronic charge governs the affinity of the CA to components of AC showcasing a charge imbalance. Recalling the composition and structure of AC, only PGs naturally show a net negative fixed charge density.

Otherwise, external alterations in the physiological environment of ECM (i.e., cleaving of negative charges in acid environment) make collagen fibrils a valuable target for the CA molecules \cite{Nemetschek1979Topochemistry}. Clinical CAs are water-soluble small molecules that can be categorized into ionic and non-ionic. The former are preferred to the aims of PG quantitation, thanks to non-covalent electrostatic Coulomb repulsion exerted from negative fixed charge density of PGs.
The efficacy of anionic CAs has been demonstrated in various AC models, particularly those involving pathological degeneration. The stronger inverse correlation with PG content justifies the broad adoption of anionic CAs compared to non-ionic CAs \cite{Jin2023Effects,Dufour2021Repair,Voert2020Contrast,Sugasawa2020Characterization,Meng2020Diffusion,Steppacher2019Ultrasonic,Michalak2019Concurrent,Raines2017Efficacy}. 
On the contrary, non-ionic CAs are limited to reveal AC morphology, as their molecules permeate the ECM according to the interstitial fluid distribution, and occasionally with the collagen content \cite{Mirahmadi2018Aging,Fan2015Correlations}. The commercial availability of clinical CAs allows their prompt adoption in CECT protocols, aside from adjustments in concentration and osmolality.
As a result, numerous works based on clinical ionic CAs extracted quantitative information from AC of different models, including large animals, rodents and human  \cite{Fleischer2023Early,Ve2021Comparison,Flynn2021Anionic,Blom2021Single,Besler2021Quantitative,Stewart2017Synthesis,Tsai2019Effects,Gatenholm2019Spatially,Reece2018Contrast,Saukko2017Dual,Mashiatulla2017Murine,Karhula2017Effects,Visser2017Groove,Tiel2016Quantitative,Stok2016Three,Mittelstaedt2016Topographical,Wang2015Assessment,Li2015Observation,Bagi2015Correlation,Renders2014Contrast,Kerckhofs2014Contrast,Fu2014Impaired,Thote2013Localized,Van2012CT,Kotwal2012Initial,Kerckhofs2012Contrast,Siebelt2011Quantifying,Siebelt2011Clinically,Xie2010Nondestructive,Bansal2010Contrast,Xie2009Quantitative,Piscaer2008In,Palmer2006Analysis,Entezari2014Effect,Wang2012Evaluation}. 
Furthemore, anionic CA-enhanced protocols served as a useful tool in monitoring the progression of osteorthritis in several preclinical studies, as confirmed by reference methods such as histology \cite{Zhang2023Cationic,Lin2023Intra,Jo2023In,Honkanen2023Dual,McKinney2022Sodium,Reece2020Reduced,Cubria2020Evaluation,
McKinney2019Therapeutic,Willett2016Quantitative,Bagi2015Effect,Siebelt2014Inhibited,Siebelt2014FK506}.
Despite the readiness of clinical CAs, the anti-correlation to PG content is biased by factors in addition to CA accumulation \cite{Taylor2009Comparison}. Moreover, anionic CAs reasonably enhance the tissue discernability at cost of conspicuous CA concentrations, with further concerns related to beam-hardening effects \cite{Michalak2019Concurrent}.

In the realm of experimental CAs, the formulation of novel molecules focused on more advantageous mechanisms of interactions with AC components. The first experimental cationic CAs \cite{Joshi2009Effect} offered higher iodine content per molecule and electrostatic attraction to PGs, yielding superior discrimination of AC. 
Further correlations with AC composition were confirmed with reference methods (i.e., histology and biochemical assay), at significantly lower concentrations compared to clinical anionic molecules \cite{Bansal2011Cationic,Bansal2011Contrast,Stewart2013Contrast,Lakin2015Contrast,Honkanen2016Cationic,Lakin2016Contrast,
Karhula2017Micro,Nelson2019Evaluation,Valerio2023Effect,Durongbhan2023microCT,Chan2023Contrast,Fantoni2022Cationic} (see Figure \ref{fig_intro_CE}a).
Furthermore, mechanical tests confirmed the intertwined relationship between the attenuation of cationic CA and the mechanical behavior of AC, considering the tissue’s elastic, frictional, and equilibrium properties \cite{Stewart2017Synthesis,Lakin2015Contrast,
Lakin2013Cationic,Nickmanesh2018Contrastenhanced,Dourthe2019Assessment,
Freedman2020dGEMRIC,Boos2021Contrast,Gao2021Influence,
Nelson2021Cationic,Nelson2021Quantitative,Nelson2024Longitudinal,
Zhang2019Protocol}.

Besides iodine, other radiopaque candidates have been selected for AC visualization. Gadolinium is a chemical element of lanthanide series, and enters the composition of MRI CAs, owing to its paramagnetic properties. In the clinical formulations, gadolinium-based CAs are predominantly non-ionic or anionic, and are employed as well as the iodine-based CAs, to signal the water content and PG distribution \cite{Taylor2009Comparison,Freedman2020dGEMRIC}. Due to its different K-edge energy (50.2 keV) with respect to iodine (33.2 keV), gadolinium-based CAs can be used simultaneously along with iodine-based CAs in DE and multi-energy protocols  \cite{Bhattarai2018Quantitative,Saukko2019Simultaneous,Bhattarai2020Dual,Honkanen2020Triple,Honkanen2020Synchrotron,Bhattarai2021Effects,Saukko2022Dualcontrast}. In the wake of novel formulations, a cationic molecule containing bismuth yielded greater correlations to reference methods for assessing PGs, compared to analogous molecules  \cite{Bhattarai2023Computed}.

Laboratory stains, usually employed in histology and transmission electron microscopy, offer a valid alternative to clinical CAs. For instance, polyoxometalates such as phosphotungstic acid (PTA) e phosphomolybdic acid (PMA) considerably increase the X-ray visibility of the exposed soft tissue \cite{Pauwels2013exploratory}. Nevertheless, different binding mechanisms oblige fixation and dehydration of the tissue, leading ot irreversible alterations of its properties. While effective for collagen quantification \cite{Ojanen2023Micro,Clark2021High,
Zhu2021Contrast,Dunham2021Increased,Ylitalo2019Quantifying,
Kwok2019Knee,Kun2017Contrast,Nieminen2015Determining,
Das2014Rapid} (see Figure \ref{fig_intro_CE}b), imaging protocols based on polyoxometalates imply alterations of the AC incompatible to mechanical testing \cite{Davis2024Comparison}. Only in recent times, the adoption of physiological environments has limited the forthcoming modifications of AC properties, following the exposure to such stains \cite{Davis2023Development}.

Further efforts to enhance the imaging outcome of existing CAs include functionalizing ioxaglate to specifically bind to PG \cite{Zhang2023Cationic} or tyrosine (an amino acid naturally containing iodine) to type-II collagen of ECM \cite{Fowkes2022Imaging}. In recent years, increasing attention has been addressed to nanoparticles (NPs) due to their small size (1–1000 nm) and tunable physicochemical properties, including shape, hindrance, and composition \cite{Freedman2014Tantalum,Hsu2020Nanoparticle}. For instance, tantalum-based NPs enable the quantitation of PG \cite{Lawson2021Tantalum,Jantti2024Cationic} or ensure the detection of defects in the tissue \cite{Omelchenko2024New}.

In existing literature, only few studies addressed CECT on chondrocyte imaging, owing to the demanding spatial resolution. Rather than the cells themselves, the imaging of lacunae (namely, the cavities in ECM enclosing one or more chondrocytes, with typical size of 10 µm \cite{Pedersen2013Comparative}) allows for more relaxed imaging protocols, carried out with nano-focus X-ray tubes. Following the contrast enhancement, lacunae were reportedly depicted as focalized sites featuring a radiopacity different from the surrounding ECM \cite{Karhula2017Micro,Nieminen20173D}. Nevertheless, laboratory-based systems equipped with X-ray tubes require long acquisition times, raising concerns about the stability of the sample over time. Such issue can be overcome with SR. For instance, the high monochromatic flux significantly lowers the scan time and ensures spatial resolution down the cellular level \cite{Danalache2021Exploration} (see Figure \ref{fig_intro_CE}c).

\subsubsection{Dual-energy imaging}

\begin{figure}[!h]
\centering\includegraphics[width=5in]{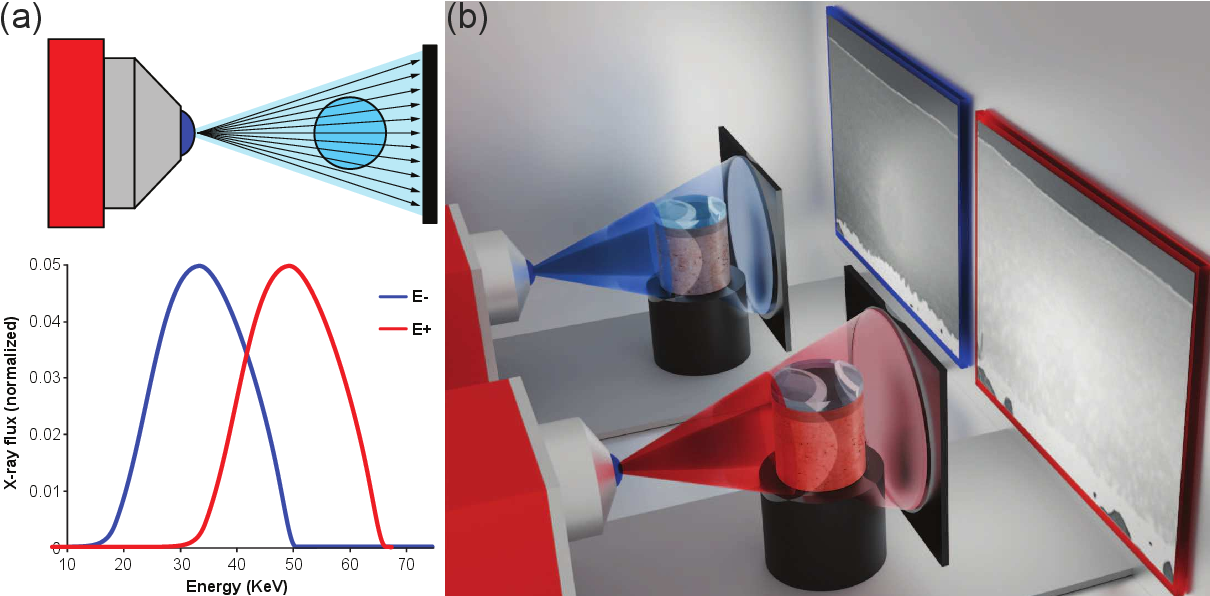}
\caption{(a) Scheme of DE technique. The procedure includes two distinct X-ray exposures, each at a different beam energy. The low-energy and high-energy components serve as input for decomposition algorithms, to deliver two material maps. (b) Reconstructed section from low-energy E- and high-energy E+ binned photons is reported on blue and red screen, respectively. In the low-energy and high-energy screens are reported sections from reconstructed volume of AC, following the immersion in a gadoteridol/CA4+ mixture. In particular, the contrast in low- and high- energy sections is due mainly to gadoteridol and CA4+, respectively. Images on screens of panel (b) adapted from the work of Honkanen et al. \cite{Honkanen2020Synchrotron}, published under Creative Commons license (CC BY 4.0).}
\label{fig_intro_DE}
\end{figure}

\begin{table}[!h]
\caption{Summary of research papers including X-ray imaging of AC with DE protocols, ordered from the most to the least recent. For each paper, the corresponding reference is reported, along with the CA used, its net electric charge $q$, the model for the AC tissue, the size of the specimen, the sourcre type, the voltage $V$ or associated energy $E$ of the X-ray beam (m. monochromatic), the acquisition time $T$ and the voxel size of the reconstructed volume $p$.}
    \centering
    \scriptsize
    \resizebox{\textwidth}{!}{
\begin{tabular}{ccccccccc}
    \hline
ref. & contrast agent & $q$ (e) & model & $S$ (mm) & source & $V$/$E$ & $T$ (min) & $p$ (µm) \\ \hline
        \cite{Saukko2022Dualcontrast} & CA4+ & +4 & equine & NA & X-ray tube & 50/90 kVp & 2.6 & 59 \\
        ~ & gadoteridol & 0 & ~ & ~ & ~ & ~ & ~ & ~ \\ \hline
        \cite{Bhattarai2021Effects} & CA4+ & +4 & human & Ø 8 & X-ray tube & 50/90 kVp & NA & 40 \\
        ~ & gadoteridol & 0 & ~ & Ø 8 & ~ & ~ & ~ & ~ \\ \hline
        \cite{Honkanen2020Triple} & CA4+ & +4 & bovine & Ø 7 & synchrotron & 32/34 keV m. & 2.15 & 6.5 \\
        ~ & gadoteridol & 0 & ~ & ~ & ~ & ~ & ~ & ~ \\
        ~ & Bi-NPs & 0 & ~ & ~ & ~ & ~ & ~ & ~ \\ \hline
        \cite{Honkanen2020Synchrotron} & CA4+ & +4 & human & Ø 8 & synchrotron & 32/34 keV m. & 2.15 & 6.5 \\
        ~ & gadoteridol & 0 & ~ & ~ & ~ & ~ & ~ & ~ \\ \hline
        \cite{Bhattarai2020Dual} & CA4+ & +4 & human & Ø 8 & X-ray tube & 50/90 kVp & NA & 40 \\
        ~ & gadoteridol & 0 & ~ & ~ & ~ & ~ & ~ & ~ \\ \hline
        \cite{Saukko2019Simultaneous} & CA4+ & +4 & bovine & Ø 4 & synchrotron & 25/37 keV & 2.1 & 6.5 \\
        ~ & gadoteridol & +4 & ~ & ~ & ~ & ~ & ~ & ~ \\ \hline
        \cite{Bhattarai2018Quantitative} & CA4+ & +4 & human & Ø 8 & X-ray tube & 50/100 kVp & 18/10 & 25 \\
        ~ & gadoteridol & 0 & ~ & ~ & ~ & ~ & ~ & ~ \\ \hline
    \end{tabular}
    }
  \label{tab:table_DE}%
  \vspace*{-4pt} 
\end{table}%

Dual-energy (DE) and, more in general, multi-energy imaging is based on several acquisitions taken at different X-ray beam energies. This method can provide information on sample composition, if unknown, or conversely, allows reconstructing the density distribution of certain elements, provided that the composition is known a-priori \cite{Alvarez1976Energy}. Dual-energy imaging is particularly useful when CAs or naturally occurring high-Z materials are present within the sample. By making use of attenuation properties specific to the material of interest, such as the K-edge, it is possible to discriminate structures containing the high-Z element, even though they have an overall attenuation similar to the surrounding features \cite{Lehmann1981Generalized}. This is achieved through dedicated material decomposition algorithms, which, having as input two (or more) attenuation images at different energies, provide density maps of two (or more) materials. A scheme of DE technique is reported in Figure \ref{fig_intro_DE}.
The ideal conditions for DE imaging are obtained using monochromatic beams, as in the case of SR \cite{Thomlinson2018K,Perion2024high} or laser-driven radiation sources \cite{Kulpe2019K}. In this case, the material decomposition problem can be solved exactly, due to the absence of beam-hardening, which is commonly encountered when polychromatic spectra are used.

In the context of AC X-ray imaging, DE technique offers its potential in discerning two or more CAs, when used simultaneously. In particular, radiopaque CAs with different attenuation signatures (i.e., K-edge energies) are selected to track single AC components (see Table \ref{tab:table_DE}). The X-ray energies are carefully chosen to enclose the attenuation discontinuities of the selected CAs separately. By applying material decomposition algorithms to DE dataset, the density maps of the elements of interest are obtained \cite{Lehmann1981Generalized}.
Such protocols were applied to understand the various factors influencing CA diffusion in AC \cite{Bhattarai2018Quantitative,Bhattarai2020Dual,Bhattarai2021Effects,
Saukko2022Dualcontrast}. Specifically, iodine accounted for the PG content, owing to the electrostatic repulsion between the anionic CA and negative fixed charge density of glycosaminoglycans. In contrast, gadolinium CA tracked the water content.

The concurrent determination of ionic and non-ionic partitions improved the correlation between the PG-related iodine distribution and AC mechanical properties \cite{Bhattarai2018Quantitative,Saukko2019Simultaneous,Honkanen2020Triple,
Bhattarai2021Effects,Honkanen2019Imaging,
Paakkari2021Quantitative}. Dual-energy methods can be further implemented with SR. The monochromaticity and high fluxes of the generated beam ensure improved image quality and reduced scan times \cite{Saukko2019Simultaneous,Honkanen2020Triple,Honkanen2020Synchrotron}.
Nevertheless, the material decomposition comes with the increase of radiation dose deposited to the sample, proportional to the number of X-ray exposures (i.e., number of X-ray energies) and the exposure time. On the other hand, the development of novel detectors has enabled the simultaneous measurement of transmitted X-ray photons in multiple energy windows, by using one single polychromatic X-ray exposure.

\subsubsection{Detector-based spectral imaging}

\begin{figure}[!h]
\centering\includegraphics[width=5in]{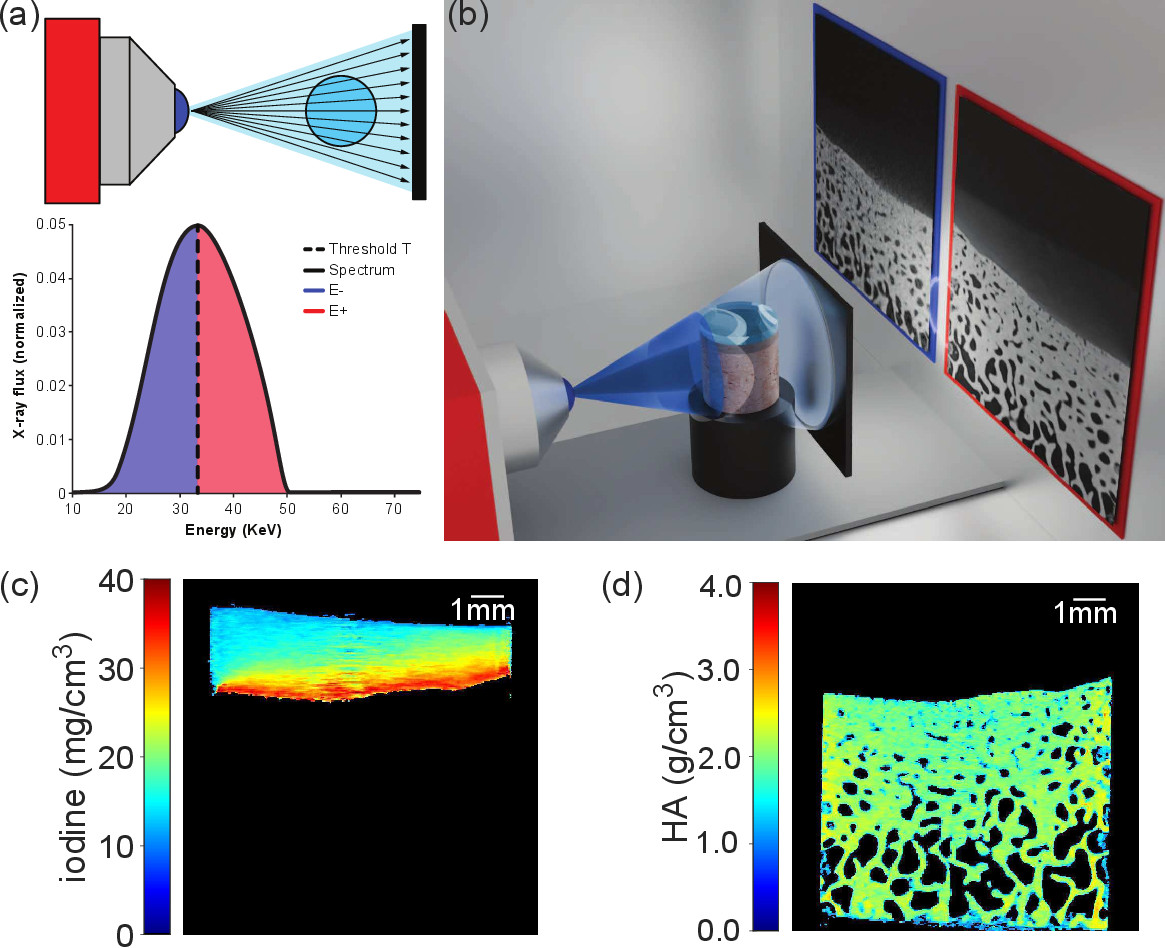}
\caption{(a) Scheme of spectral imaging technique. The procedure includes a single X-ray exposure. The discrimination occurs on-chip, following the selection of an energy threshold. According to the latter, photons are binned in low-energy and high-energy components. Analogously to DE technique, both components are required for decomposition algorithms to provide two material maps. (b) Reconstructed section from low-energy E- and high-energy E+ binned photons of a bovine osteochondral sample is reported on blue and red screen, respectively. (c, d) Section of density maps, as a result of the decomposition algorithm. (c) Iodine density map accounting for iodine-based CA4+ diffused in AC. (d) Hydroxyapatite (HA) density, accounting for calcium-rich bone tissue. Images in panels (b), (c) and (d) adapted from the work of Fantoni et al. \cite{Fantoni2024Quantitative}, published under Creative Commons license (CC BY 4.0).}
\label{fig_intro_spectral}
\end{figure}

Photon-counting detectors discriminate the incoming photons, depending on their energy and recently have found applications in several imaging fields \cite{Flohr2020Photon}. Photon-counting technology is based on the direct conversion of incoming photons by semiconductors sensors and the following discrimination of the signal at the pixel level. Unlike scintillator elements in energy-integrating detectors, which achieve X-ray conversion into visible light, semiconductors enable the direct conversion of incoming photons into electric signal, with an intensity proportional to the energy deposited by each interacting photon to the sensitive region of the detector \cite{Taguchi2013Vision}. Within each pixel of a PCD, this signal is discriminated based on one (or more) energy-calibrated threshold. If multiple energy thresholds are available, photons can be detected and grouped over multiple energy intervals (or bins). In this way, a single X-ray exposure yields multiple attenuation maps, as many as the energy bins defined by the thresholds. The application of decomposition algorithms selectively enhances or eliminates structures in the sample, similarly to the DE imaging \cite{Rajendran2023Improved}. The principles of spectral imaging are summarized schematically in Figure \ref{fig_intro_spectral}. Unlike DE methods, polychromatic X-ray beams are employed, and no specific constraints on X-ray spectrum are required. Therefore, the implementation of PCDs is suitable for clinical \cite{Rajendran2022First} and laboratory \cite{Brombal2023PEPI} X-ray systems.

Detector-based spectral imaging yielded promising results in terms of AC discernability, although the spatial resolution of clinical scanners remains limited \cite{Luetkens2023Ultra,Kim2023Computed}. This issue is partially solved by laboratory-based spectral systems equipped with small pixel ($<$ 100 µm) PCDs, where the spatial resolution has been significantly improved. Previous experiments verified the suitability of spectral imaging ex-vivo and in-vitro on human specimens, exposed to several CAs \cite{Paakkari2021Quantitative,Paakkari2023Tantalum,Tuppurainen2024Revealing}.  Alternatively, other works exploited a single CA but aimed to distinguish between the contrast-enhanced AC and the underlying bone tissue \cite{Rajendran2017Quantitative,Baer2021Spectral,Fantoni2024Quantitative} (see Figures \ref{fig_intro_spectral}c and \ref{fig_intro_spectral}d), even down to the smallest pixel size  reported for a spectral imaging system, namely 34 µm (see Table \ref{tab:table_PCD}).

\begin{table}[!h]
\caption{Summary of research papers including X-ray imaging of AC with PCD-based spectral protocols, ordered from the most to the least recent. For each paper, the corresponding reference is reported, along with the CA used, its net electric charge $q$, the model for the AC tissue, the size of the specimen, the PCD model, the voltage $V$ of the X-ray beam, the acquisition time $T$ and the voxel size of the reconstructed volume $p$.}
    \centering
    \small
        \resizebox{\textwidth}{!}{
    \begin{tabular}{ccccccccc}
    \hline
        ref. & contrast agent & $q$ (e) & model & $S$ (mm) & PCD & $V$ & $T$ (min) & $p$ (µm) \\ \hline
        \cite{Tuppurainen2024Revealing} & \ce{Ta2O5} NPs & +4 & equine & Ø 8.5 & Xcounter & 150 kVp & NA & 68 \\
        ~ & iodixanol & 0 & ~ & ~ & XC-Flite FX15 & 120 kVp & ~ & ~ \\ \hline
        \cite{Fantoni2024Quantitative} & CA4+ & +4 & bovine & Ø 10 & Pixirad1-PixieIII & 50 kVp & 120 & 34 \\ \hline
        \cite{Paakkari2021Quantitative} & CA4+ & +4 & human & Ø 8 & 0.75mm CdTe & 100 kVp & NA & 87 \\
        ~ & gadoteridol & 0 & ~ & ~ & 100um pixel size & ~ & ~ & ~ \\ \hline
        \cite{Baer2021Spectral} & ioxaglate & -1 & bovine & Ø 8 & 2-mm CdTe & 80/120 kVp & NA & 110 \\
        ~ & gadobenate & -3 & human & ~ & Medipix3RX & ~ & ~ & ~ \\ \hline
        \cite{Rajendran2017Quantitative} & ioxaglate & -1 & human & 25 & 2-mm CdTe & 80 kVp & NA & 73 \\
        ~ & ~ & ~ & ~ & ~ & Medipix3RX & ~ & ~ & ~ \\ \hline
    \end{tabular}
    }
  \label{tab:table_PCD}%
\end{table}%

\subsection{Refraction-based X-ray Imaging Techniques: Phase-Contrast}
Phase-contrast imaging exploits the phase shift of X-rays interacting with the sample. Phase-contrast rises from interference phenomena involving the wavefront distorted by the sample \cite{Zhou2008Development}. The modulation of the wavefront can be described by the phase shift accumulated by the incoming X-ray wave while traversing the sample, which is proportional to the decrement from unity of the real part $\delta$ of the complex refractive index. Conversely, it can be demonstrated that X-ray attenuation, which is used in conventional X-ray imaging, is proportional to the imaginary part $\beta$  of the complex refractive index ($n$ = 1 - i $\beta$ - $\delta$). In the case of soft tissues and energies used in conventional radiology (10 – 100 keV), PC is advantageous as $\delta$ is two to three orders of magnitude larger than $\beta$, allowing to highlight internal structures related to subtle fluctuations of density or interfaces without using exogenous CAs \cite{Momose2000Blood}. The sample-induced phase-shift can be related to refraction effects where local modifications to the wave vector can be described, in a simplified ray-tracing model, as small-angle deviations (order of microradians) of the impinging X-rays \cite{Peterzol2005effects}.
As the X-ray wave phase cannot be directly measured with imaging detectors, several PC techniques have been developed to transform the phase shift into detectable intensity modulations. Some of these techniques pose strict requirements on the X-ray beam coherence (namely monochromaticity and small focal spot size), and are therefore primarily implemented within SR facilities \cite{Cloetens1996Phase,Momose1996Phasecontrast,Snigirev1995On}. More recently, also alternative PC setups with relaxed coherence requirements have been introduced, therefore making PC imaging accessible within laboratory settings \cite{Pfeiffer2006Phase,Olivo2007coded}. In the following subsections the image formation principles of the most widely used PC techniques will be illustrated.

\subsubsection{Propagation-based phase-contrast}

\begin{figure}[!h]
\centering\includegraphics[width=5in]{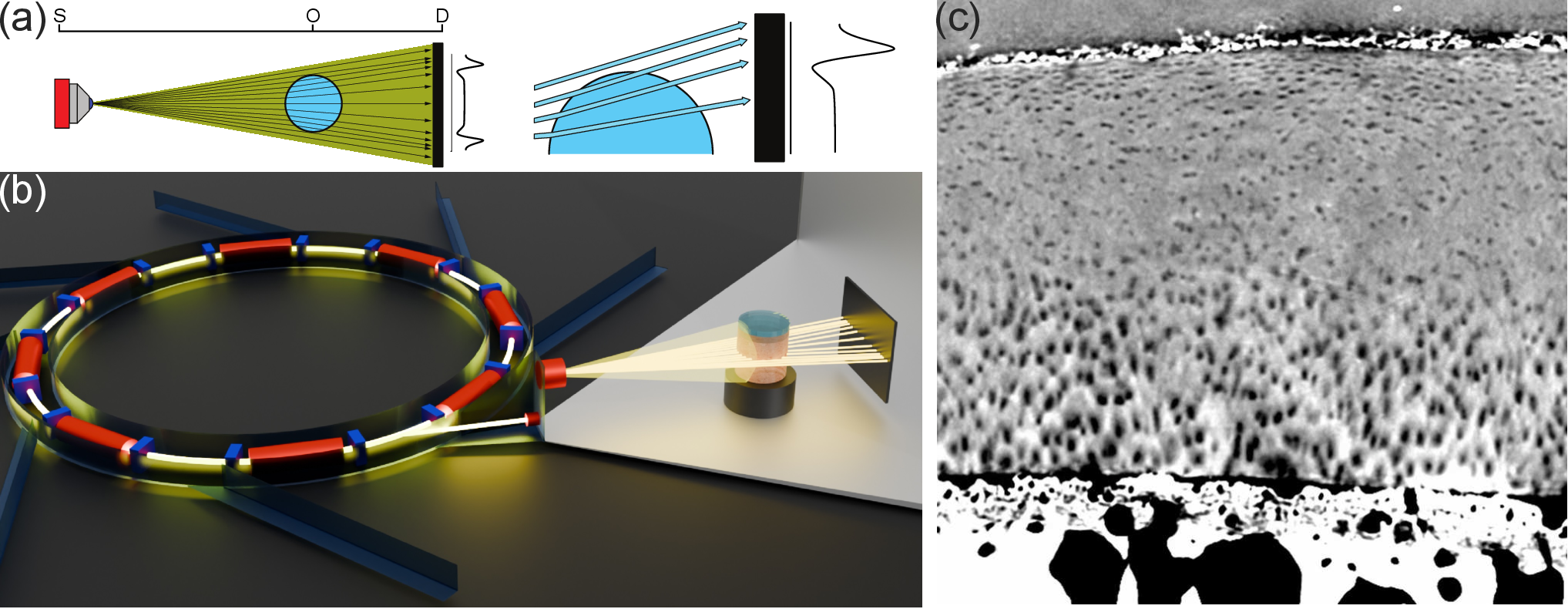}
\caption{(a) Scheme of PBPC. Unlike absorption modalities, PBPC bases its mechanism on the distance between the sample and the detector. In particular, such distance is selected to produce the edge-enhancement, originating from interfaces between different materials. (b) Three-dimensional rendering of a typical PBPC setup. The edge-enhancement requires a high degree of spatial coherence (i.e., SR). (c) Section from PBPC volume of healthy human AC. Besides the interface AC-air on the top, darker spots in the tissue, attributable to chondrocytes' lacunae, are clearly recognizable. Image adapted from the work of Horng et al. \cite{Horng2021Multiscale}, published under Creative Commons license (CC BY 4.0).}
\label{fig_intro_PC}
\end{figure}

Propagation-based phase-contrast (PBPC) imaging is the simplest PC technique to implement, as no optical element is required \cite{Wilkins1996Phase}. PBPC requires increasing sample-to-detector distance to generate a phase shift-induced interference pattern on the detector. This is generally visible as a couple of dark/bright fringes corresponding to interfaces between materials with different refractive index, and it is referred to as edge-enhancement effect.

Since the spatial coherence of the X-ray beam is crucial for PBPC \cite{Wilkins1996Phase}, this technique can be implemented with SR, laser-driven systems, or high-end microfocus X-ray tubes \cite{Ikeura2008In} (see Table \ref{tab:table_PBPC}). A scheme of image formation via PBPC is displayed in Figure \ref{fig_intro_PC}. Images featuring the edge-enhancement effect can be further processed with phase-retrieval algorithms, which enable the recovery of the phase-shift information \cite{Brombal2018Phase} and enhance the signal-to-noise ratio, hence the visibility, of soft tissue structures \cite{Gureyev2017On,Brombal2018Phase}.

\begin{table}[H]
\caption{Summary of research papers including X-ray imaging of AC with PBPC protocols, ordered from the most to the least recent. For each paper, the corresponding reference is reported, along with the model for the AC tissue, the size of the specimen $S$, the source type, the voltage $V$ or associated energy $E$ of the X-ray beam (m. monochromatic, p. polychromatic, p.b. pink beam), the acquisition time $T$, the voxel size of the reconstructed volume $p$ and the object-to-distance ODD.}
    \centering
    \small
	\resizebox{\textwidth}{!}{
    \begin{tabular}{cccccccc}
    \hline
        ref. & model & $S$ (mm) & source & $V$/$E$ & $T$ (min) & $p$ (µm) & ODD (cm) \\ \hline
        \cite{Dejea2024In} & bovine & Ø 4 & synchrotron & 21keV m. & 0.08-0.66 & 2.75 & 40 \\ \hline
        \cite{Bissardon2022In} & bovine & Ø 4-6 & synchrotron & 17 keV m. & 66 & 3.5 & 250 \\ \hline
        \cite{Horng2021Multiscale} & human & Ø 7 & synchrotron & 60keV m. & NA & 6.1 & 1100 \\
        ~ & ~ & ~ & ~ & 55keV p. & ~ & 0.7 & 120 \\
        ~ & ~ & ~ & ~ & 17keV m. & ~ & 0.325-0.1 & 10 \\ \hline
        \cite{Broche2021Calcified} & murine & NA & synchrotron & 52 keV m. & 8 &  6.1; 6.06 & 1100 \\
        ~ & ~ & NA & X-ray tube & 60 kVp & NA & ~ & ~ \\ \hline
        \cite{Tozzi2020Full} & bovine & Ø 3-4 & X-ray tube & 40 kVp & NA & 2.02-2.56 & NA \\ \hline
        \cite{Rack2020Tomo} & human & Ø 7 & synchrotron & 26.1 keV p.b. & NA & 5.1 & 500 \\ \hline
        \cite{Madi2020In} & murine & NA & synchrotron & 12-25 keV p.b. & 1.1-7.3 & 0.8-1.6 & NA \\
        ~ & ~ & ~ & ~ & 19 keV, m. & 37.5 & 1.1 & NA \\ \hline
        \cite{Geith2018Quantitative} & human & NA & synchrotron & 60 keV m. & 1.5 & 46 & 700 \\ \hline
        \cite{Yoon2015Phase} & murine & NA & synchrotron & NA & NA & NA & NA \\ \hline
        \cite{Sun2015Phase} & rabbit & NA & synchrotron & 5-18 keV p. & NA & 10 & NA \\ \hline
        \cite{Li2014Quantitative} & murine & NA & synchrotron & 14 keV m. & NA & 3.7 & NA \\ \hline
        \cite{Horng2014Cartilage} & human & NA & synchrotron & 60 keV m. & 90 & 46 & 700 \\ \hline
        \cite{Zehbe2012Imaging} & bovine & Ø 2 & synchrotron & 14 keV m. & 60-180 & 1.752 & 0.5-30.0 \\ \hline
        \cite{Lee2010Articular} & murine & NA & X-ray tube & 80 kVp & 30 & 9/6.6 & 32.2 \\ \hline
        \cite{Ismail2010X} & horse & 0.2-1 & X-ray tube & 70 kVp & NA & 13 & NA \\
        ~ & human & ~ & ~ & 11.2 keV, m. & ~ & 15 & NA \\ \hline
        \cite{Choi2010In} & murine & NA & synchrotron & 7-14 keV p. & NA & 1.48 & 5 \\ \hline
        \cite{Zehbe2010Going} & bovine & NA & synchrotron & 10-15 keV m. & 40 & 1.6 & 15 \\ \hline
        \cite{Shimao2005Refraction} & human & 20 & synchrotron & 30 keV m. & NA & NA & NA \\ \hline
    \end{tabular}
    }
  \label{tab:table_PBPC}%
\end{table}%

In the literature, PBPC has been demonstrated as a suitable imaging technique, not only in visualizing AC without the need for CA, but also in unveiling the most subtle structures. The ability to clearly distinguish AC was confirmed by studies based on planar \cite{Shimao2005Refraction,Ismail2010X,Choi2010In} and tomosynthesis approach \cite{Sun2015Phase,Yoon2015Phase}. Reduced thickness and augmented roughness of AC surface have been recognized as hallmarks of AC degeneration in diseased models \cite{Yoon2015Phase,Geith2018Quantitative}. The transfer of PBPC to tomographic acquisitions confirmed these observations \cite{Lee2010Articular,Li2014Quantitative,Broche2021Calcified}. Boosting the resolution of the acquisition system down to few µm, the sensitivity of PBPC technique to cellular pattern was further enhanced with specific CAs and  validated by independent methods such as histology \cite{Clark2020Propagation,Ruan2013Quantitative,Clark2020Exploratory} (see Table \ref{tab:table_CEPBPC}). Besides the coherence of the beam, the object-to-detector distance is crucial for the visualization of lacunae, namely the interfaces between the inner chondrocytes from the surrounding ECM.

\begin{table}[!h]
\caption{Summary of research papers including X-ray imaging of AC with contrast-enhanced PBPC protocols, ordered from the most to the least recent. For each paper, the corresponding reference is reported, along with the contrast agent and its charge $q$, the model for the AC tissue, the size of the specimen, the source type, the voltage $V$ of the X-ray beam, the acquisition time $T$, the voxel size of the reconstructed volume $p$ and the object-to-distance ODD.}
    \centering
    \small
    \resizebox{\textwidth}{!}{
    \begin{tabular}{ccccccccc}
    \hline
        ref. & contrast agent & $q$ (e) & model & $S$ (mm) & $V$ & $T$ (min) & $s$ (µm) & ODD (cm) \\ \hline
        \cite{Jayaram2020Leukocyte} & RHT & NA & murine & NA & NA & NA & NA & NA \\
        ~ & cacodylic acid & NA & ~ & ~ & ~ & ~ & ~ & ~ \\
        ~ & \ce{OsO4} & NA & ~ & ~ & ~ & ~ & ~ & ~ \\ \hline
        \cite{Clark2020Exploratory} & PTA & -3 & ovine & Ø10 & 80 kVp & NA & 4.5 & 9.5 \\
        ~ & ~ & ~ & ~ & ~ & 60 kVp & NA & 10.2 & 17.2 \\ \hline
        \cite{Clark2020Propagation} & PTA & -3 & bovine & Ø 3 & 40 kVp & NA & 1.97-2.85 & 4.0-55.8 \\ \hline
        \cite{Stone2019Combinatorial} & PTA & -3 & murine & NA & NA & NA & NA & NA \\ \hline
        \cite{Nixon2018DiseaseModifying} & \ce{OsO4} & NA & murine & NA & NA & NA & NA & NA \\
        ~ & ~ & ~ & equine & ~ & ~ & ~ & ~ & ~ \\ \hline
        \cite{Ruan2013Quantitative} & \ce{OsO4} & NA & murine & NA & 40-80 kVp & 65-1067 & 0.54-10 & 2.5-7.5 \\ \hline
        \cite{Ruan2013Proteoglycan} & \ce{OsO4} & NA & murine & NA & 40 kVp & NA & 4 & 7.5 \\ \hline
        \cite{Ruan2013Pain} & RHT & NA & murine & NA & NA & NA & NA & NA \\
        ~ & cacodylic acid & NA & ~ & ~ & ~ & ~ & ~ & ~ \\
        ~ & \ce{OsO4} & NA & ~ & ~ & ~ & ~ & ~ & ~ \\ \hline
	\end{tabular}
	}
  \label{tab:table_CEPBPC}%
\end{table}%

Larger sample-to-detector distances and higher beam energies of the X-ray beam were demonstrated to be beneficial for the optimal visualization of structural details \cite{Zehbe2010Going}. Noteworthily, the proposed imaging protocol distinguishes also chondrocyte from the surrounding lacunae, as confirmed by scanning electron microscope images \cite{Zehbe2010Going,Zehbe2012Imaging}. Horng et al. further pushed the evaluation of healthy and diseased AC down at sub-cellular level, highlighting even the cellular nucleus and bundles of collagen fibers with the highest spatial resolution of 0.1 µm \cite{Horng2021Multiscale} (see Figure \ref{fig_intro_PC}c).

Other studies, rather than focusing on cellular-scale details, analyzed the morphology of AC on larger scales. The acquisition of whole cadaveric human joints yielded not only the delineation of several soft tissues, but also the differentiation of AC layers. The latter feature was observed to vary substantially from healthy to osteoarthritic AC. The authors speculated that the loss of chondrocytes associated to the pathology could imply a variation in the electron density associated with the retrieved signal \cite{Horng2014Cartilage}.

The quantitative analysis was also carried out with the application of Paganin phase-retrieval, assigning different gray level windows, according to the heterogeneous arrangement of components in ECM \cite{Horng2021Multiscale,Bissardon2022In,Rack2020Tomo}. Unless more complex acquisition schemes are implemented, such as in the case of holotomography \cite{Cloetens1999Holotomography}, a quantitatively accurate phase-retrieval requires strict assumptions on sample composition \cite{Paganin2002Simultaneous}. Alternatively, the phase information can be accessed with different imaging configurations, making use of optical elements in the imaging setup. These approaches will be discussed in the following subsections.

\subsubsection{Analyzer-based imaging}

\begin{figure}[!h]
\centering\includegraphics[width=5in]{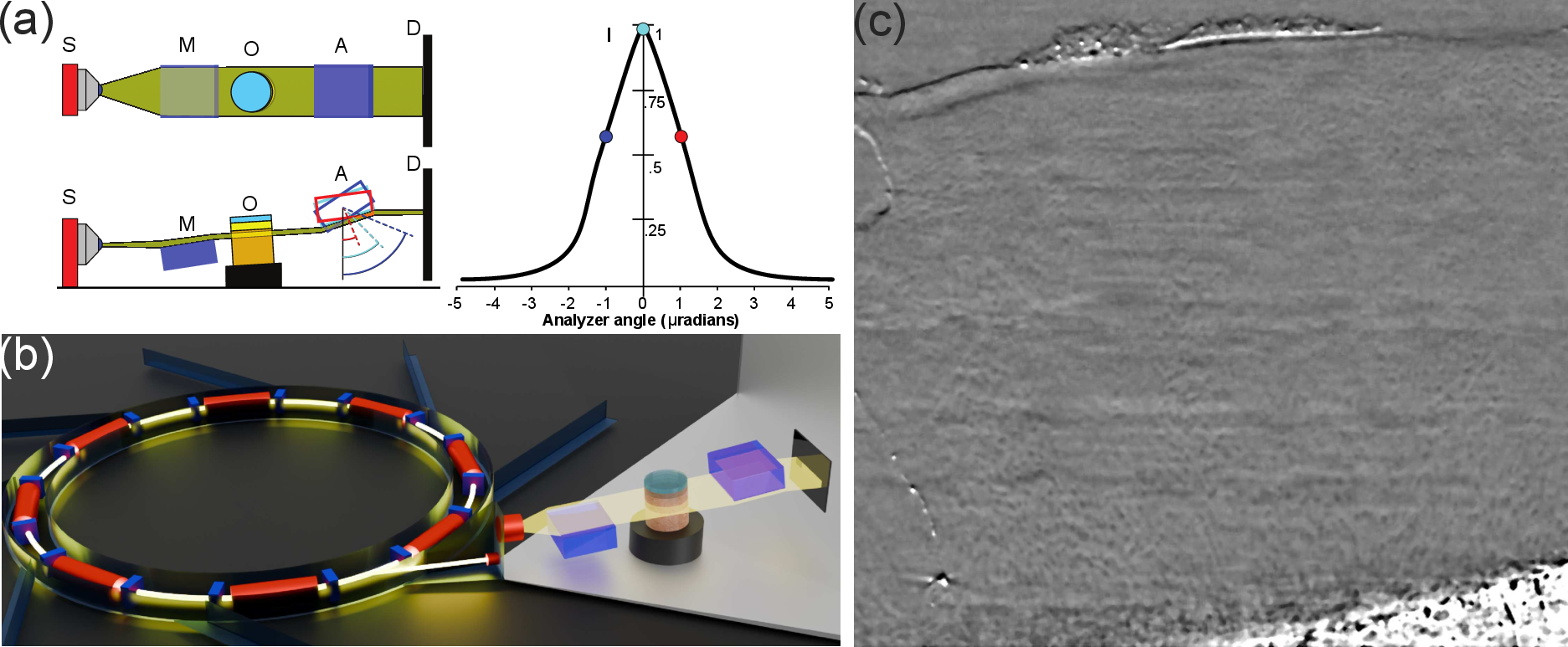}
\caption{(a) Scheme of ABI. The insertion of the analyzer crystal A acts as an angular filter on the photons transmitted by the sample O. Only photons satisfying the angular acceptance, determined by the rocking curve (in the right bottom), are transmitted by the analyzer A and reach the detector D. (b) Three-dimensional rendering of a typical ABI setup. The ABI requires monochromatic and laminar X-ray beam, typically provided by synchrotron facilities. (c) Section from reconstructed ABI volume of healthy human AC. Margins of AC are highlighted, as well as the speckle pattern conferred by the distribution of chondrocytes' lacunae. Image adapted from the woek of Nagarajan et al. \cite{Nagarajan2015Integrating}, published under Creative Commons license (CC BY 4.0).}
\label{fig_intro_ABI}
\end{figure}

Analyzer-based Imaging (ABI) makes use of a perfect crystal, referred to as analyzer, set between the object and the detector \cite{Ingal1995X,Davis1995Phase,Chapman1997Diffraction}, and requires a monochromatic and laminar X-ray beam. The analyzer crystal acts as an angular filter of the incoming X-ray beam. When the analyzer crystal is set to an angle that satisfies the Bragg’s condition with respect to the incoming beam, X-rays are transmitted (i.e., diffracted). The transmission efficiency decreases as the analyzer is rotated off the Bragg’s condition. The curve describing the X-ray transmission as a function of the analyzer crystal’s angle is called rocking curve \cite{Ingal1995X}.

As the angular acceptance of the rocking curve is in the order of some microradians, the analyzer crystal allows the translation of X-ray refraction into intensity modulation with high sensitivity. By acquiring images at different angles of the analyzer crystal, that is, different points on the rocking curve, both attenuation and (differential) phase signal can be obtained via dedicated algorithms. Additionally, if at least three images are acquired in different positions of the rocking curve, ABI allows the extraction of the scattering or dark-field (DF) signal, hence giving access to structures below the system’s spatial resolution \cite{Davis1995Phase,Chapman1997Diffraction,Talbot1836LXXVI}.
While ABI is usually associated with high image quality and quantitative accuracy, its use is mostly limited within SR facilities due to the need for intense monochromatic laminar beams. The mechanism behind the image formation of ABI is outlined in Figure \ref{fig_intro_ABI}.

The potential of ABI for AC depiction was explored in the first 2000s at different synchrotron facilities. Although the first images were radiographies impressed on radiographic film, they provided robust results in terms of discernability of soft tissues \cite{Crittell2007Diffraction,
Li2013Assessment,Muehleman2006Preliminary,Miki2023Visualization,
Wagner2005Chance,Li2005Reliability}. The positioning of the analyzer crystal was crucial for AC visualization. Depending on the analyzer tilting, only those photons emerging from the sample within the angular acceptance range of the rocking curve (order of few µrad) reach the image plane. Therefore, different structures in AC are highlighted at different angular positions of the analyzer, i.e. at different working positions on the rocking curve \cite{Mollenhauer2002Diffraction,Li2003Radiography,
Muehleman2003Radiography,Majumdar2004Diffraction,Muehleman2004Yes,
Shimao2004Application} (see Table \ref{tab:table_ABI}). For example, if the peak of the rocking curve is chosen, only photons experiencing no or little deflection traversing the sample are transmitted by the analyzer, ensuring an optimal scatter rejection. Conversely, by setting the orientation of the analyzer crystal in one slope of the rocking curve, the system becomes sensitive to refracted photons, which are transmitted with a higher (lower) probability if the refraction angle is towards the top (tail) of the rocking curve \cite{Mollenhauer2002Diffraction,Li2003Radiography,
Muehleman2003Radiography,Shimao2004Application}. The overall effect of the analyzer orientation on AC radiographs is the edge enhancement of tissue boundaries and of highly oriented structures such as collagen, delineated as bright or dark fringes. Similarly, defects within AC are noticeable, and rendered differently depending on the analyzer angle \cite{Mollenhauer2002Diffraction,Wagner2005Options}.

With the advent of digital detectors, ABI shifted toward tomographic applications, enabling three-dimensional reconstructions of joints from various models, at superior spatial resolutions compared to CT and MRI \cite{Rhoades2015Diffraction,Gasilov2016Hard,Coan2010In}. The efficacy of ABI was also assessed on whole intact human knee joints, despite the prolonged exposure time and limited vertical aperture of the X-ray beam \cite{Li2009Phase}. Aside from the clear distinction of macroscopic structures, ABI is also suitable for the imaging of AC microarchitecture. The high sensitivity to subtle changes in the refraction index allows for the depiction of collagen arcades \cite{Muehleman2004X} and lacunae \cite{Issever2008Analyser}, down to the spatial resolution of few µm.
High-resolution ABI further enabled cellular characterization of healthy and osteoarthritic AC, distinguishing chondrocyte alignment, zonal distribution, and fibrillation\cite{Coan2010Characterization,Nagarajan2015Volumetric,Nagarajan2015Integrating,Abidin2018Deep} (see Figure \ref{fig_intro_ABI}c). Different algorithms of phase extraction from ABI images were also studied, and their performance was compared on a common AC sample. Weighting the advantages and pitfalls of the single approaches, those yielding the independent reconstruction of apparent absorption, refraction and scattering signals resulted in the best visualization \cite{Diemoz2010Comparison,Diemoz2010Absorption}.

Interestingly, several studies exploited ABI for its multimodal potential. Rather than the refraction component alone, Muehleman and colleagues applied the method known as multiple-image radiography (MIR) to the visualization of AC. MIR consists in the acquisition of multiple images (namely, more than three) at several positions of the rocking curve. By using eleven different crystal orientations, the researchers explored the suitability of MIR in detecting USAXS signal with high precision. Notably, the scattering image distinguished the connective tissues, owing to their inner collagen arrangement \cite{Muehleman2006Multiple}. The latter approach was later extended to conventional and limited-angle tomography techniques \cite{Majidi2014Limited}.

Only few studies implemented ABI with table-top systems. The requisite of monochromaticity is satisfied also by kilovoltage tubes with the adoption of monochromator crystals, although the X-ray beam flux results severely reduced \cite{Muehleman2010In}. The refraction component clearly depicts AC from other soft tissues in intact joints, and distinguishes AC samples at different degradation stages, with results comparable to histology \cite{Muehleman2009Diffraction,Fogarty2011In}. Nonetheless, the implementation of ABI to table-top systems remains the most challenging compared to other PC techniques.

\begin{table}[H]
\caption{Summary of research papers including X-ray imaging of AC with ABI protocols, ordered from the most to the least recent. For each paper, the corresponding reference is reported, along with the model for the AC tissue, the size of the specimen, the source type, the energy $E$ of the X-ray beam (m. monochromatic, q.m. quasi-monochromatic), the number of analyzer positions $N_{A}$, the acquisition time $T$ and the voxel size of the reconstructed volume $p$.}
    \centering
    \small
    \resizebox{\textwidth}{!}{
    \begin{tabular}{cccccccc}
    \hline
        ref. & model & size (mm) & source & $E$ & $N_{A}$ & $T$ (min) & $p$ (µm) \\ \hline
        \cite{Miki2023Visualization} & rat & NA & synchrotron & 20-35 keV m. & NA & NA & 9 \\ \hline
        \cite{Abidin2018Deep} & human & NA & synchrotron & 26 keV q.m. & NA & NA & 8 \\ \hline
        \cite{Gasilov2016Hard} & rabbit & NA & synchrotron & 51 keV m. & 2 & 10-60 & 46 \\ \hline
        \cite{Rhoades2015Diffraction} & swine & NA & synchrotron & 40 keV m. & 5 & NA & 18.7 \\ \hline
        \cite{Nagarajan2015Volumetric,Nagarajan2015Integrating} & human & Ø 7 & synchrotron & 26 keV q.m. & NA & NA & 8 \\ \hline
        \cite{Majidi2014Limited} & human & NA & synchrotron & 40 keV & 25 & NA & 50 \\ \hline
        \cite{Li2013Assessment} & human & NA & synchrotron & 40 keV m. & 1 & 5 & 31.2 \\ \hline
        \cite{Fogarty2011In} & equine & 80x35x30 & X-ray tube & 22.15 keV m. & 17 & NA & 160; 153 \\ \hline
        \cite{Muehleman2010In} & human & NA & X-ray tube & 22.15 keV m. & 15 & 240 & 160; 153 \\ \hline
        \cite{Diemoz2010Absorption} & human & Ø 7 & synchrotron & 26 keV m. & 5 & NA & 16 \\ \hline
        \cite{Diemoz2010Comparison} & human & Ø 8 & synchrotron & 25 keV m. & 5 & NA & 16 \\ \hline
        \cite{Coan2010In} & guinea pig & NA & synchrotron & 52 keV m. & NA & NA & 47 \\ \hline
        \cite{Coan2010Characterization} & human & Ø 7 & synchrotron & 26 keV m. & 1 & NA & 8 \\ \hline
        \cite{Muehleman2009Diffraction} & human & NA & X-ray tube & 59 keV m. & NA & NA & 50 \\ \hline
        \cite{Li2009Phase} & human & NA & synchrotron & 51 keV m. & 1 & 1380 & 56.2 \\ \hline
        \cite{Issever2008Analyser} & human & Ø 2.6 & synchrotron & 20 keV m. & 1 & 7000 & 3.6 \\ \hline
        \cite{Crittell2007Diffraction} & murine & NA & synchrotron & 15 keV m. & 2 & NA & NA \\ \hline
        \cite{Muehleman2006Preliminary} & canine & NA & synchrotron & 40 keV m. & 3 & NA & 50 \\ \hline
        \cite{Muehleman2006Multiple} & human & NA & synchrotron & 40 keV m. & 11 & NA & NA \\ \hline
        \cite{Wagner2005Chance} & human & NA & synchrotron & NA & 5 & NA & NA \\ \hline
        \cite{Wagner2005Options} & human  & NA & synchrotron & 20-50 keV m. & 1-3 & NA & NA \\ \hline
        \cite{Li2005Reliability} & human & NA & synchrotron & 30 keV m. & 3 & 0.2 & 50 \\ \hline
        \cite{Shimao2004Application} & human & 20 & synchrotron & 15 keV m. & NA & NA & 50 \\ \hline
        \cite{Muehleman2004X} & human & 40x60, Ø15 & synchrotron & 17/25 keV m. & NA & NA & 5 \\ \hline
        \cite{Muehleman2004Yes} & human & NA & synchrotron & 40 keV m. & 2 & NA & 50 \\ \hline
        \cite{Majumdar2004Diffraction} & human & Ø 15 & synchrotron & 17 keV m. & NA & NA & NA \\ \hline
        \cite{Muehleman2003Radiography} & rabbit & NA & synchrotron & 30 keV m. & 3 & NA & NA \\ \hline
        \cite{Li2003Radiography} & human & NA & synchrotron & 40 keV m. & 5 & 10; 0.5 & 50-75 \\ \hline
        \cite{Mollenhauer2002Diffraction} & human & NA & synchrotron & 18/30 keV m. & NA & 0.07-0.1 & 50 \\ \hline
    \end{tabular}
    }
  \label{tab:table_ABI}%
\end{table}%

\subsubsection{Gratings interferometry}

\begin{figure}[!h]
\centering\includegraphics[width=5in]{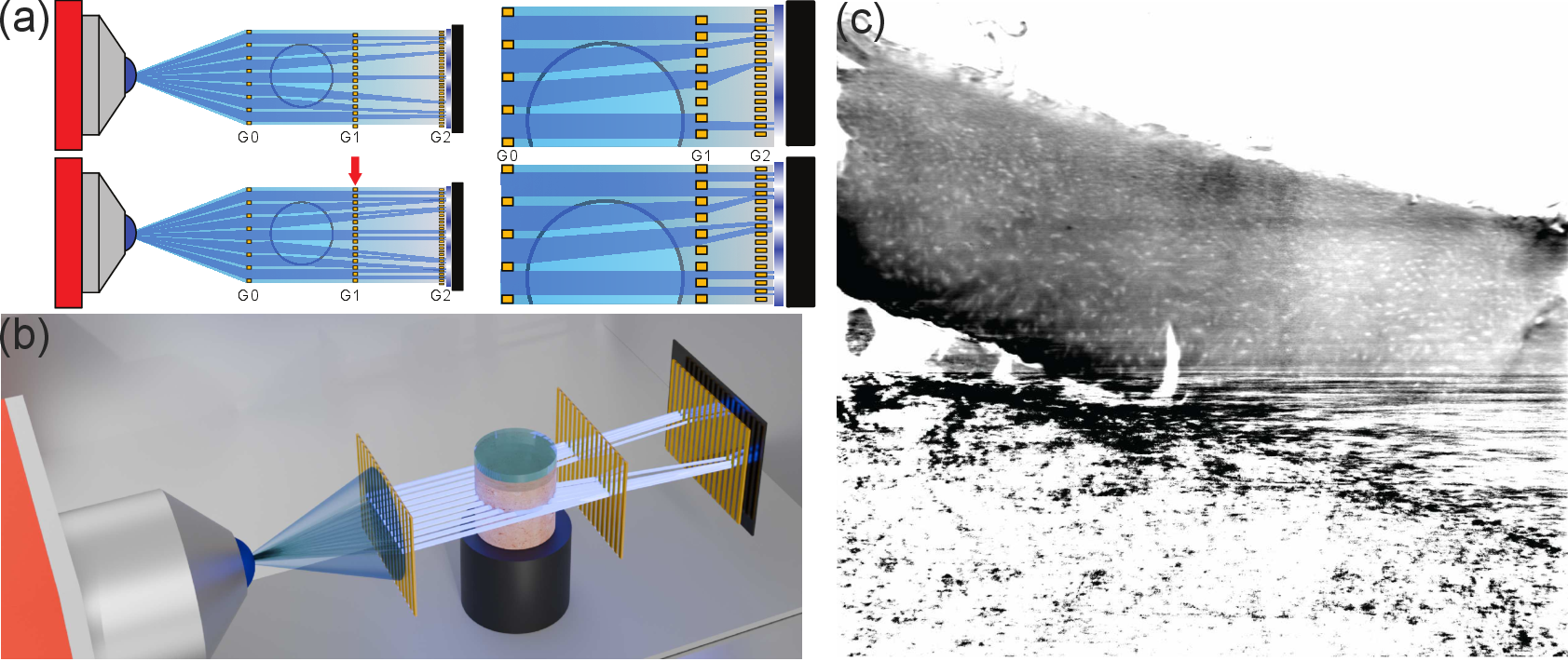}
\caption{(a) Scheme of GI, in Talbot-Lau configuration. Provided the spatial coherence of the X-ray beam, only the sample gratings G1 and the detector gratings G2 are required. Otherwise, source gratings G0 are required, as in the depicted case. The distance of the detector gratings is determined according to the occurrence of the Talbot self-image effect. The lateral shift of the detector gratings allows to collect the refraction and the scattering signals. (b) Three-dimensional rendering of a typical Talbot-Lau GI setup, featuring a conventional X-ray source (e.g., X-ray tube). (c) Section from reconstructed GI volume of healthy human AC. Within AC, spots featuring higher intensity are distinguishable, attributable to chondrocytes' lacunae. Image adapted from the work of Schulz et al. \cite{Schulz2017Multimodal}, published under Creative Commons license (CC BY 3.0).}
\label{fig_intro_GI}
\end{figure}

Gratings interferometry (GI) is based on the Talbot self-image effect produced by regular periodic structures with a period in the order of a few micrometers, referred to as gratings, set along the direction of propagation of X-rays. Due to Fresnel diffraction, diffraction patterns with the same periodicity of the grating are formed at specific distances from the grating, which are equal to multiples of the Talbot distance \cite{Talbot1836LXXVI}. The first GI studies exploited the spatially-coherent synchrotron X-ray beam and made use of two gratings, referred to as G1 and G2, generally positioned downstream from the sample \cite{David2002Differential,Momose2003Demonstration}. The grating G1, referred to as phase-grating, is set close to the sample and it is made ideally of X-ray transparent materials, introducing a periodic phase-shift that is responsible for the Talbot diffraction pattern. The grating G2, referred to as absorption grating, is positioned at a multiple of the Talbot distance and acts as an analyzer, partially absorbing the X-rays. As a function of the relative displacement or misalignment of the two gratings, a modulation in the X-ray intensity recorded at the imaging plane is observed. Similarly to ABI, when a sample is introduced in the beam, the recorded intensity is modified according to sample’s attenuation, phase, and scattering properties, which can be retrieved though dedicated algorithm \cite{Brombal2023Hybrid}.
In the wake of these SR-based results, Pfeiffer and colleagues successfully adapted GI to conventional low-brilliance X-ray sources \cite{Pfeiffer2006Phase}. To overcome the smearing of the diffraction pattern that would occur using a source with limited spatial coherence, a third absorption grating close to the source, is inserted in the so-called Talbot-Lau configuration. The source grating G0 generates a periodic array of repeated sharp line sources which are individually coherent despite being globally non-coherent. This arrangement produces a diffraction pattern which, as in the Talbot case, can be transformed into intensity modulations at the detector plane by displacing G1 and G2 \cite{Pfeiffer2008Hard}. The Talbot-Lau scheme for GI is displayed in Figure \ref{fig_intro_GI}. 

\begin{table}[!h]
\caption{Summary of research papers including X-ray imaging of AC with GI protocols, ordered from the most to the least recent. For each paper, the corresponding reference is reported, along with the model for the AC tissue, the size of the specimen $S$, the source type (Mo molybdenum, W tungsten), the focal spot size $s_{FS}$, the voltage $V$ or associated energy $E$ of the X-ray beam (m. monochromatic), the number of grating positions $N_{G}$, the acquisition time $T$ and the voxel size of the reconstructed volume $p$.}
    \centering
    \small
    \resizebox{\textwidth}{!}{
\begin{tabular}{ccccccccc}
    \hline
        ref. & model & $S$ (mm) & source & $s_{FS}$ (µm) & $V$/$E$ & $N_{G}$ & $T$ (min) & $p$ (µm) \\ \hline
        \cite{Kawano2022In} & porcine & Ø 3 & synchrotron & NA & 20 keV m. & 5 & 20 & 4.46 \\ \hline
        \cite{Yoshioka2020Imaging} & human & NA & X-ray tube (W) & 400 & 40 kVp & NA & 0.32 & 85 \\ \hline
        \cite{Herzen20193D} & bovine & Ø 6 & X-ray tube (Mo) & 400 & 40 kVp & 11 & NA & 41 \\ \hline
        \cite{Khimchenko2018Implementation} & human & Ø 5.4 & X-ray tube & 0.9-2.7 & 42 kVp & 7 & NA & 23.3 \\
        ~ & ~ & ~ & synchrotron & ~ & 19 keV & NA & 2.3 & 23.3 \\
        ~ & ~ & ~ & synchrotron & ~ & 52 keV & NA & 5.1 & 23.3 \\ \hline
        \cite{Schulz2017Multimodal} & human & Ø 5 & synchrotron & NA & 19 keV & 5 & NA & 2.3 \\
        ~ & ~ & Ø 5 & X-ray tube & NA & 40 kVp & 5 & NA & 3 \\
        ~ & ~ & entire knee & X-ray tube & NA & 180 kVp & ~ & 1020 & 65 \\ \hline
        \cite{Schulz2016Hierarchical} & human & Ø 5 & X-ray tube & NA & 52 keV & 4 & NA & 5.1 \\
        ~ & ~ & Ø 10 & synchrotron & NA & 19 keV & 5 & NA & 2.3 \\ \hline
        \cite{Nagashima2014Application} & human & NA & X-ray tube (W) & 300 & 40 kVp & 5 & 0.25 & 85 \\ \hline
        \cite{Momose2014X} & human & NA & X-ray tube (W) & 450 & 40 kVp & 3 & 19 & 85 \\ \hline
        \cite{Tanaka2013Cadaveric} & human & NA & X-ray tube (W) & 400 & 40 kVp & 3 & 19 & 90 \\ \hline
        \cite{Tada2012Fabrication} & porcine & NA & X-ray tube (W) & 300 & 49 kVp & NA & NA & NA \\ \hline
        \cite{Murakoshi2012Feasibility} & porcine & NA & X-ray tube (W) & 300 & 49 kVp & NA & NA & NA \\ \hline
        \cite{Kiyohara2012Development} & human & NA & X-ray tube (W) & 300 & 40kVp & 5 & NA & 85 \\ \hline
        \cite{Itoh2011Two} & chicken & NA & synchrotron & NA & 17.5 keV m. & NA & 0.008 & 12 \\
        ~ & ~ & ~ & ~ & ~ & 35 keV m. & ~ & ~ & ~ \\ \hline
        \cite{Diemoz2011simplified} & human & Ø 8 & synchrotron & NA & 32 keV m. & 3 & NA & 16 \\ \hline
        \cite{Stutman2011Talbot} & human & NA & X-ray tube (W) & 60 & 40 kVp & NA & NA & 50 \\
        ~ & ~ & ~ & ~ & 300 & 40 kVp & 16 & NA & 50 \\ \hline
        \cite{Makifuchi2010Development} & chicken & NA & X-ray tube (W) & 300 & 40 kVp & NA & NA & NA \\ \hline
        \cite{Kido2010Bone} & chicken & NA & X-ray tube (W) & 300 & 40 kVp & NA & NA & NA \\ \hline
        \cite{Momose2009Grating} & chicken & NA & X-ray tube (W) & 300 & 40 kVp & NA & 0.67 & 18 \\ \hline
    \end{tabular}
    }
  \label{tab:label_GI}%
\end{table}%

Since its inception, GI has been applied to the visualization of articular soft tissues. Synchrotron-radiation implementations of GI allowed a good delineation of the edges of AC \cite{Itoh2011Two,Murakoshi2012Feasibility,Tada2012Fabrication}. Similarly, laboratory-based GI setups using polychromatic and non-pointlike focal spot sources has delivered differential PC images enabling the visualization of AC \cite{Momose2009Grating,Kido2010Bone,Makifuchi2010Development} (see Table \ref{tab:label_GI}). Advances in grating fabrication promoted the translation of GI to clinical settings, with ex-vivo studies on human metacarpophalangeal joints highlighting AC margins \cite{Kiyohara2012Development,Tanaka2013Cadaveric,Nagashima2014Application,Stutman2011Talbot}, followed by optimized low-dose in-vivo imaging of healthy volunteers \cite{Tanaka2013Cadaveric,Nagashima2014Application,Stutman2011Talbot,Momose2014X}.
More recently, a conventional X-ray tube-driven GI system has been tested for the assessment of rheumatoid arthritis, to ascertain its sensitivity to tissue alterations. Independent evaluations, namely MRI and clinical scores, have supported the results, suggesting that the GI apparatus distinguishes healthy AC from pathological tissue \cite{Yoshioka2020Imaging}. Albeit the promising results in terms of AC visibility, the works reported until now have focused on GI to radiography, as the tomographic implementation would raise further concerns on deposited dose.

In the research frame, the translation of GI to tomography delivers three-dimensional maps of absorption, refraction and scattering (or DF) components \cite{Diemoz2011simplified,Khimchenko2018Implementation}. The refraction-enhanced highlighted AC microstructures such as chondrocyte clusters, while energy-dependent protocols enabled simultaneous imaging of mineralized and soft tissues \cite{Schulz2016Hierarchical,Schulz2017Multimodal} (see Figure \ref{fig_intro_GI}c).

Grating interferometry imaging also reportedly revealed differences in collagen content between superficial and deep AC layers. These variations reflect the gradient of electron density related to collagen content, making GI a quantitative tool complementary to state-of-the-art MRI maps. Interestingly, GI imaging is not prone to biases induced by variation in water content, unlike MRI. In this aspect, GI could provide sound information of ECM integrity, regardless of the interstitial fluid \cite{Herzen20193D}.

\subsubsection{Edge-illumination imaging}
\begin{figure}[!h]
\centering\includegraphics[width=5in]{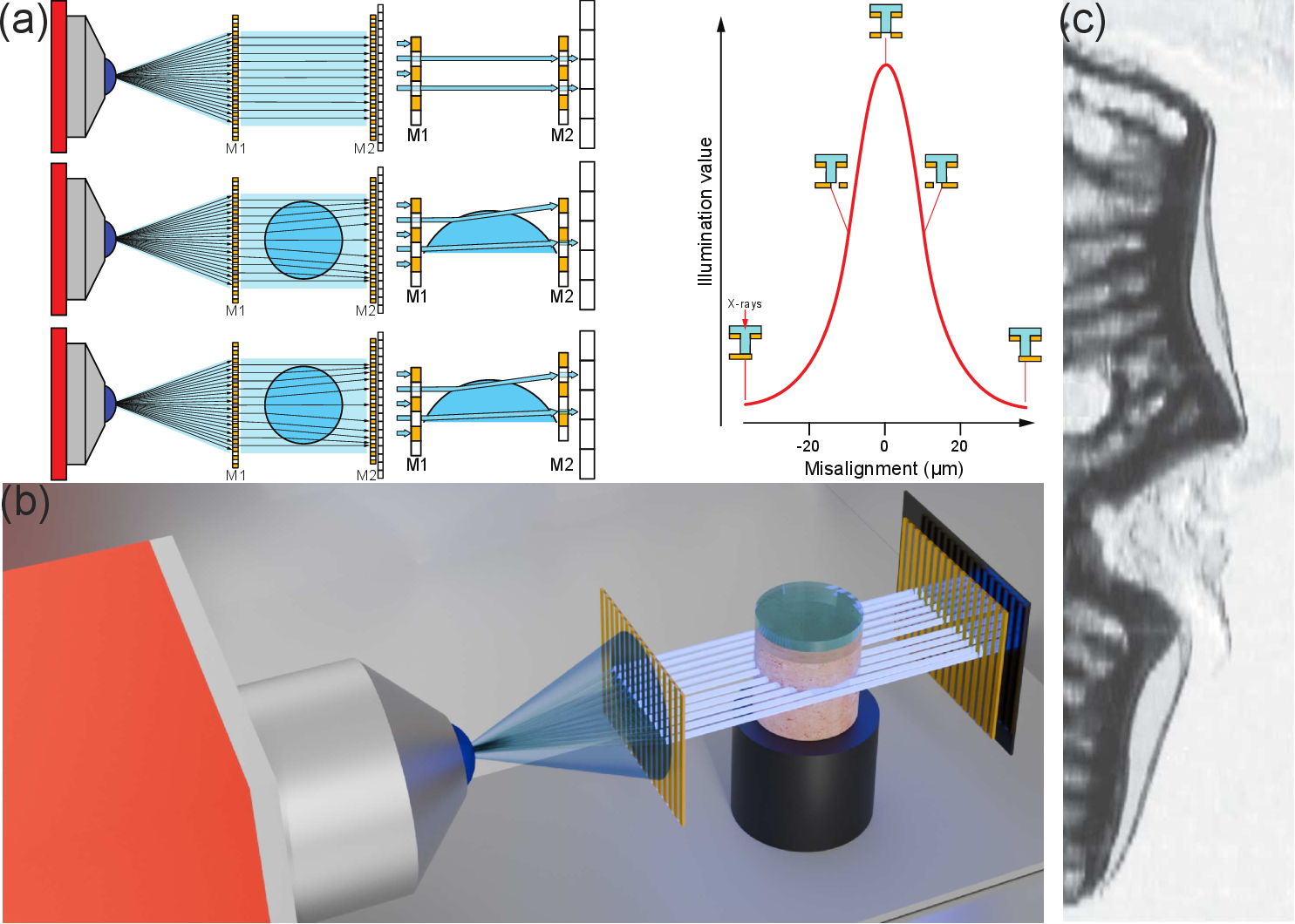}
\caption{(a) Scheme of EI imaging. When illuminated, the sample mask M1 and the detector mask M2 produce an illumination pattern, replicating the periodicity of the masks. The overall effect of masks’ lateral displacements on the intensity modulation of a single pixel is summarized by the illumination curve (in the bottom right). The insertion of the sample modifies the illumination pattern, according to interactions of different nature. In particular, the absorption, the refraction and the scattering of photons either attenuate, deviate or broaden the beamlets, and alter the modulation of the illumination curve. The lateral shift of the detector mask M2 allows the retrieval of those signals, as different sensitive areas of detector are differentially exposed. (b) Three-dimensional rendering of a typical EI setup. As reported in literature, nearly all experiments featured conventional X-ray sources (e.g., X-ray tubes). (c) Planar image of 1 mm-section from healthy murine osteochondral tissue. Besides the depiction of mineralized tissues (black structures), the interface of AC is clearly recognizable. Image adapted from the work of Marenzana et al. \cite{Marenzana2014Synchrotron}, published under Creative Commons license (CC BY 3.0).}
\label{fig_intro_EI}
\end{figure}

Edge illumination (EI) imaging allows access to phase information by means of regular absorbing structures (i.e. masks) with a period in the order of tens of micrometers, positioned along the X-ray propagation direction \cite{Olivo2007coded}. A typical EI setup features two masks, positioned upstream of the sample (M1) and close to the detector (M2), respectively. Each mask is structured as a linear array of absorbing material, such as gold, interleaved with X-ray transparent septa, or apertures, with a dimension of a few micrometers. Both masks share the same periodicity and aperture size, apart from the geometrical magnification factor depending on their relative position with respect to the X-ray source. The masks’ period matches the detector pixel pitch, such that each periodic feature of the mask corresponds to one detector pixel column (or row). 
The role of the mask M1 is to structure the incoming X-ray beam into a series of independent, non-interfering narrow beams, or beamlets. In the absence of a sample, when the apertures of mask M2 are aligned with the apertures of M1, the beamlets are fully transmitted to the detector. If M1 and M2 are laterally displaced, the transmission decreases (ideally) reaching zero if the apertures are completely misaligned. The curve describing pixel-by-pixel the X-ray transmission across the two masks as a function of their lateral displacement is known as the illumination curve and plays a role conceptually analogous to the rocking curve in ABI. When a sample is introduced, the illumination curve is modulated in terms of area, due to X-ray absorption, lateral position, due to refraction, and width, due to scattering. A scheme of EI principles of image formation is shown in Figure \ref{fig_intro_EI}. By using suitable retrieval algorithms and acquiring images at (at least) three positions on the illumination curve, all these effects can be uncoupled producing attenuation, phase and DF maps of the sample \cite{Endrizzi2014Hard}.
Although GI and EI share some similarities, EI is a non-interferometric technique as the beamlets are sufficiently spaced to be individually analyzed, hence not relying on the formation of diffraction patterns. As for the Talbot-Lau GI, the EI configuration does not require strict constraints on X-ray source coherence and it is implemented with commercial, polychromatic X-ray sources \cite{Brombal2023PEPI,Endrizzi2014Hard}.

\begin{table}[!h]
\caption{Summary of research papers including X-ray imaging of AC with EI protocols, ordered from the most to the least recent. For each paper, the corresponding reference is reported, along with the model for the AC tissue, the size of the specimen, the source type, the voltage $V$ or associated energy $E$ of the X-ray beam (m. monochromatic), the number of mask positions $N_{M}$ the acquisition time $T$, the voxel size of the reconstructed volume $p$ and the object-to-distance ODD.}
    \centering
    \small
    \resizebox{\textwidth}{!}{
    \begin{tabular}{cccccccc}
    \hline
        ref. & model & $S$ (mm) & source & $V$/$E$& $N_{M}$ & $T$ (min) & $p$ (µm) \\ \hline
        \cite{Massimi2022Replacing} & chicken & NA & X-ray tube & 40 kVp & NA & 0.03 & NA \\ \hline
        \cite{Marenzana2014Synchrotron} & murine & NA & X-ray tube & 40 kVp & NA & NA & 9 \\
        ~ & ~ & NA & synchrotron & 17 keV m. & NA & NA & 5 \\ \hline
        \cite{Endrizzi2013Edge} & rat & NA & X-ray tube & NA & NA & NA & NA \\ \hline
        \cite{Marenzana2012Visualization} & rat & NA & X-ray tube & 40 kVp & 4 & 1 & 20 \\ \hline
    \end{tabular}
    }
  \label{tab:table_EI}%
\end{table}%

In the context of AC imaging, EI has been mostly implemented within table-top systems (see Table \ref{tab:table_EI}). Notably, differently from all other mentioned techniques, the spatial resolution of an EI system is not related to the size of the X-ray focal spot or to the detector pixel size, but it is ultimately determined by the aperture size of the pre-sample mask, allowing an additional degree of flexibility in trading-off acquisition time and image quality \cite{Olivo2007coded,Diemoz2015Angular,Olivo2001innovative}. 
The research group at University College London, who pioneered EI, demonstrated the viability of EI implemented with uncollimated and non-microfocus X-ray tube, for the visualization of small lesions in rat AC \cite{Marenzana2012Visualization}. Despite the relatively thin AC layer, EI images provided its visualization in both air and aqueous environments. The measurements of tissue thickness were comparable to CE-based studies and gold standard techniques, such as histology \cite{Endrizzi2013Edge,Marenzana2014Synchrotron}. The yield in refraction images are comparable to synchrotron-based ABI system and confirms the soundness of using the refraction signal to highlight the AC layer \cite{Marenzana2014Synchrotron} (see Figure \ref{fig_intro_EI}c).
One further implementation of EI has considered the insertion of a structured detector to replace the detector mask (M2), demonstrating a clear visualization of AC in refraction image, similarly to conventional EI setup \cite{Massimi2022Replacing}.

\subsection{Scattering-based X-ray imaging: dark-field imaging}
Dark-field imaging conveys contrast from ultra small-angle X-ray scattering (USAXS) photons, which arise from the interaction of primary X-rays with microscopic structures in the sample. These structures are typically not resolved with conventional attenuation-based techniques, owing to insufficient spatial resolution or to the low radiopacity of the structures \cite{Pfeiffer2008Hard}. 
Under the ray tracing approximation, USAXS can be understood as the effect of multiple refraction causing locally a diffusion of the X-ray beam. The amount of diffusion can be traced back to quantities such as the average size of the scatterers or the scattering power of the sample, hence giving access to information at a spatial scale below the system’s spatial resolution \cite{Pfeiffer2008Hard,Kunisada2008X,Zanette2012Trimodal,Rigon2007Three}. As anticipated in the previous subsections, many PC methods including ABI, GI, and EI give access to the scattering signal, generally at the cost of additional exposures at different positions of the optical element.

\begin{table}[!h]
\caption{Summary of research papers including X-ray imaging of AC with DF protocols, ordered from the most to the least recent. For each paper, the corresponding reference is reported, along with the model for the AC tissue, the size of the specimen $S$, the source type, the size of the focal spot $s_{FS}$ (Cu copper), the voltage $V$ or associated energy $E$ of the X-ray beam (m. monochromatic, p.b. pink beam), the acquisition time $T$, the voxel size of the reconstructed volume $p$ and the object-to-distance ODD. *detector pitch size}
    \centering
    \small
    \resizebox{\textwidth}{!}{
    \begin{tabular}{cccccccc}
    \hline
        ref. & model & $S$ (mm) & source & $s_{FS}$ (µm) & $V$/$E$ & $T$ (min) & $p$ (µm) \\ \hline
        \cite{Esposito2023Laboratory} & rabbit & 1 & X-ray tube (Cu) & 350 & 8 keV m. & NA & 1.1* \\ \hline
        \cite{Esposito2023Technical} & equine & 2x0.5x0.5 & synchrotron & NA & 25 keV p.b. & 5 & 0.8 \\
        ~ & ~ & ~ & X-ray tube (Cu) & 350 & 8 keV m. & NA & 1.1* \\ \hline
        \cite{Ando2016Dark} & human & NA & synchrotron & NA & 20 keV m. & NA & 7.4 \\
        ~ & ~ & ~ & ~ & ~ & 12 keV m. & NA & 16 \\ \hline
        \cite{Kunisada2008X} & human & NA & synchrotron & NA & 35 keV m. & NA & 10 \\ \hline
        \cite{Shimao2007Refraction} & human & NA & synchrotron & NA & 36 keV m. & 0.92 & NA \\ \hline
        \cite{Shimao2006Articular} & human & NA & synchrotron & NA & 36 keV m. & NA & NA \\ \hline
        \cite{Shimao2005X} & human & NA & synchrotron & NA & 34.8 keV m. & NA & NA \\
        ~ & ~ & ~ & ~ & ~ & 33 keV m. & NA & NA \\ \hline
        \cite{Shimao2005Evaluation} & human & NA & synchrotron & NA & 34.8 keV m. & NA & NA \\ \hline
        \cite{Ando2004Construction} & human & NA & synchrotron & NA & 35 keV m. & NA & 10 \\ \hline
        \cite{Shimao2003Imaging} & human & NA & synchrotron & NA & 15 keV m. & NA & NA \\ \hline
    \end{tabular}
    }
  \label{tab:table_DF}%
\end{table}%

Dark-field has been explored as a complementary signal for assessing soft tissues, including AC, alongside PC refraction-based methods. Since the first studies, results demonstrated the ability of DF in depicting features of AC at a sub-pixel spatial resolution. Early applications included film-based DF radiographies acquired with the ABI technique, which highlighted the soft tissue components of excised human joints in different environmental conditions \cite{Shimao2003Imaging,Ando2004Construction,Kunisada2008X,
Ando2016Dark}. The same approach was followed also on intact joints, where the differentiation of AC structures resulted optimal, given a careful selection of orientations of the analyzer crystal \cite{Shimao2005X,Shimao2005Evaluation}. In particular, different orientations conferred major contrast to AC contour or bulk \cite{Shimao2006Articular}. Notwithstanding the promising results in radiography, the rather large number of frames required at different orientations of the analyzer significantly increases the radiation dose and the exposure time, thus limiting the translation of ABI-based DF imaging to SR-based CT (see Table \ref{tab:table_DF}). On the other hand, tomosynthesis approach requires fewer projection images, compared to CT, while giving a certain level of in-depth information. Notably, past works delivered DF tomosyntheses with doses equivalent to a single DF radiograph \cite{Shimao2007Refraction}.
Recently, the feasibility of DF has been demonstrated on alternative experimental designs. Specifically, thanks to the beam-tracking technique, that is an EI variant making use of a high-resolution detector instead of the M2 mask \cite{Vittoria2015Beam}, DF, refraction and attenuation images of AC could be obtained simultaneously \cite{Herzen20193D}. From these images, authors speculated the visualization of collagen bundles. Further radiographs performed with the same imaging configuration properly visualized lacunae of AC \cite{Esposito2023Laboratory}. Despite the two-dimensional nature, the cellular pattern was compatible to reference images from synchrotron-driven PBPC method and histology \cite{Esposito2023Technical}.

\section{Towards the in-situ evaluation of articular cartilage}

Originally conceived for trabecular bone characterization \cite{Bay1995Texture}, DVC has been applied to various musculoskeletal tissues \cite{Karali2021Micromechanical,Costa2017Micro,Disney2019Synchrotron,Pierantoni2025Quantification}. Its principles include the tracking recognizable voxel-intensity variations through sequential compression steps, the evaluation of displacement vectors, and the derivation of the strain tensor field \cite{Bay1999Digital}. In the case of AC, time-resolved rheologic experiments require multiple rapid tomographic scans, yielding speckled patterns obtained under different compressive steps. To the aims of DVC, the lacunae hosting chondrocytes serve as discrete patterns ideal for strain measurements.
Such patterns can be highlighted with several approaches. In absorption mode, CAs such as the polyoxometalates enhance the contrast between the lacunae and the surrounding ECM \cite{Clark2021High}, though fixation and staining alter the mechanical properties of the tissue. Alternatively, staining protocols could lead to the preservation of the original mechanical behavior of AC \cite{Davis2023Development,Davis2024Comparison}. Furthermore, DNA-binding stains enhance lacunae visibility \cite{Danalache2021Exploration}, but require complex processing procedures and a severe downsizing of the sample.
The combination of PC with CAs was also investigated to highlight features of AC and compensate the limited flux and coherence of laboratory microCT systems \cite{Ruan2013Quantitative,Jayaram2020Leukocyte,Stone2019Combinatorial,Nixon2018DiseaseModifying,Ruan2013Proteoglycan,
Ruan2013Pain,Clark2020Propagation,Clark2020Exploratory} (see Table \ref{tab:table_CEPBPC}).
However, the limited X-ray flux implies prolonged scan times, with significant impact on AC properties and sample stability \cite{Tozzi2020Full}. Synchrotron-radiation microCT overcomes these issues thanks to the availability of high flux X-ray beams.

For instance, a methodology to visualize and quantify cellular structures in compressed AC was recently proposed, along with a dedicated SR beamline \cite{Rack2020Tomo}. In the same beamline, the compressive method with DVC was also implemented on murine model, and yielded the visualization of hierarchical changes in AC structures. The high spatial resolution reached (0.8 µm) allowed the calculation of displacements with accuracies below 100 nm in knees of healthy and osteoarthritic mice \cite{Madi2020In}.

In addition to PBPC, the investigation of osteochondral samples under compression was accomplished with other PC methods. At the SPring-8 synchrotron facility in Japan, GI-microCT successfully resolved the cellular pattern of porcine AC, as confirmed by histology. The deformation of the cellular pattern served as input for DVC analysis. Remarkably, the density map of sample was shown to change accordingly to the imparted compression \cite{Kawano2022In}.

Despite the positive outcome of the above-described implementations, most studies making use of PC techniques require a scanning or stepping of optical elements. As a result, they are focused on imaging of AC in static configuration, because of the longer times related to acquisition procedures and the limited time-resolving capabilities of laboratory-based systems, associated to the limited X-ray flux. Interestingly, the time-resolved approach was successfully implemented with PBPC, exploiting monochromatic SR. Its implementation was demonstrated at TOMCAT beamline, at Paul Scherrer Institute in Switzerland, where a rheometric setup allowed the evaluation of dynamic behavior through time of unprocessed bovine AC samples \cite{Dejea2024In}. The tomographic experiment took advantage of the shortest exposure time to date for the dynamic investigation of connective tissues, resulting in acquisition times ranging from 40 down to 5 seconds. The advantageous experimental condition achieved spatial resolutions compatible with the size of lacunae (2.75 µm), with nearly no radiation damage induced to the tissue. Notwithstanding the detrimental laminarization of the monochromatized beam, the experiment benefited from a field of view large enough to accommodate the whole AC (5.54x3.85 mm against 4-mm diameter samples).
According to existing literature, PBPC is the most straight-forward implementation for rheometric measurements. It offers favorable acquisition time and image quality, while keeping as low as possible the complexity of the experimental setup.

\section{Benefits, challanges and future perspectives}

The above discussed X-ray imaging techniques achieve optimal results in terms of contrast and detail discernability of AC, in a variety of models and experimental conditions. Nevertheless, the choice of the imaging approach should be tailored to the rationale of the investigation. The relatively large number of studies applying CAs reflects the straightforward implementation of CECT to AC imaging. The use of ionic CAs allows for a quantitative analysis of several components of AC (see Table \ref{tab:table_CE}). Nevertheless, their possible impact on tissue’s mechanical properties draws concerns in rheologic applications. For instance, hyperosmolar CA solutions modifies tissue stiffness \cite{Turunen2012Hyperosmolaric,Pouran2016Isolated,Zimmerman2021Direct}. Furthermore, certain staining protocols alter the mechanical response of the tissue and introduce significant bias in rheometric evaluation of the whole osteochondral unit \cite{Davis2024Comparison}.

Additionally, the composition of CAs should be chosen to optimize the imaging outcome. The imaging protocol should include X-ray energies compatible with the absorption signature of the selected radiopaque element. In the context of spectral protocols, this aspect becomes more crucial. In fact, mixtures of CAs with different absorbing features facilitate the material discrimination \cite{Paterno2020Dual}. Furthermore, the CA concentration should be carefully chosen to simultaneously yield the optimal contrast among details and reduce imaging artifacts. Despite its readiness, the use of CAs with conventional microCT scasnners could impair the segmentation of tissues with similar radiopacity, as in the case of contrast-enhanced AC and mineralized tissues \cite{Xie2009Quantitative}.

Spectral approaches such as DE technique could circumvent the issue, yet facing several limitations. First, multiple acquisitions require volume co-registration in space or in time, whether images are collected with two sources at different spatial locations simultaneously, or with a single source at different time points, respectively. Second, the span of materials is restricted by the limited flexibility of spectra provided by conventional X-ray sources. Third, the material decomposition worsens if the number of materials entering the composition of the sample exceeds the beam energies \cite{Paterno2020Dual}. Last, radiation dose increases with the number of energy exposures.
Photon-counting detectors overcome these limitations, although their larger pixel sizes and/or smaller field-of-view limit the spatial resolution. The development of the latest PCD technologies (i.e., modular structure and advanced signal clustering) allows larger field of view, and a resolution finer than the pixel pitch \cite{Procz2019X,Ramilli2017Measurements}.

Phase-contrast techniques enable the visualization of several structures in AC (namely collagen arcades, lacunae and even chondrocytes) \cite{Zehbe2010Going,Zehbe2012Imaging,Horng2021Multiscale,Coan2010Characterization,Abidin2018Deep,
Schulz2016Hierarchical,Schulz2017Multimodal,Kawano2022In,Dejea2024In} without exogenous CAs, though most approaches require high-brilliance sources, ad-hoc optics, or their combination. 
Propagation-based PC is the most promising method for rheologic evaluation of AC, provided the implementation with SR \cite{Dejea2024In}. The same considerations apply to ABI, plus the additional limitations related to the use of monochromatic X-ray beams, laminar geometry and the need for multiple scans at different analyzer orientations.
Similarly to ABI, GI is a multimodal approach sensitive to refraction and scattering signals, delivering promising results in-vivo \cite{Yoshioka2020Imaging}. In research-oriented applications, GI allows for the visualization of cellular patterns in AC, pushing its applications to in-situ testing of AC \cite{Kawano2022In}. Nonetheless, high-end applications of GI make the implementation with SR essential, to compensate for the prolonged acquisition times associated with multiple-frame grating stepping.
The rigorous requirements on X-ray beam decay for non-interferometric EI, as it is easily implemented to compact laboratory-based systems. Furthermore, EI potentially retrieves phase information with only two images, by illuminating the pixels at their opposite sides, compared to GI method \cite{Munro2013quantitative}. Nevertheless, studies adopting EI for AC imaging are the minority, with no reported experience on in-situ evaluations.
Focusing on the scattering signal, DF imaging is still under development, specifically for AC evaluation. Despite the unique nature of the signal, DF imaging has been mainly limited to plain radiographs or tomosynthesis only, as the tomographic approach would significantly increase the acquisition times and the dose fractions.

Aside from in-depth analysis of the single techniques, more general observations arise by considering the examined literature. In general, high resolution is obtained at the cost of reduced sample size, increased photon fluxes and greater radiation doses. Furthermore, the evaluation of mechanical response requires volumes representative of the whole tissue (i.e., ECM), rather than single structures (i.e., collagen bundles). Recently, an imaging protocol has been proposed as optimal trade-off between the sample size and spatial resolution for the visualization of lacunae, in the context of time-resolved in-situ imaging \cite{Dejea2024In}.

Another concern arises from the time of exposure to radiation. The scanning time of each technique is largely dependent on the X-ray source characteristics, the presence of optical elements and the geometrical configuration. Synchrotron-radiation generally ensures the fastest exposure times, especially when the full spectrum (i.e., white beam) is used coupled to the PBPC technique. Conversely, the installation of any optical element in the beam generally leads to longer exposure times both at synchrotrons and laboratory-based systems.

Radiation dose remains a major concern. Planar and tomosynthesis studies report mGy-level exposures compatible with clinical translation \cite{Horng2014Cartilage,Li2009Phase,Kiyohara2012Development,Momose2014X,Ando2016Dark}, while tomographic experiments may reach several Gy, inducing macroscopic (namely, heating of the sample)\cite{Clark2021High,Cicek2016Effect} and microscopic (namely, degradation of ECM, oxidative stress, and cellular degeneration) effects \cite{Clark2020Propagation,Honkanen2020Triple,Honkanen2020Synchrotron,Horng2014Cartilage,Li2009Phase,Dejea2024In} (see Table S4 in Supplementary Material for dose values). Furthemore, the evaluation of the whole osteochondral unit should consider significant effect of prolonged X-ray exposures on bone tissues \cite{Pena2018Effect,Barth2011Characterization}. Therefore, any rheologic experiment should consider cryogenic expedients compensating the progressive dose deposition, with special mention to SR studies.

Aside from rheologic applications, more relaxed experimental conditions for AC imaging meet the use of multimodal laboratory-based X-ray systems. Preliminary experiences retrieved successfully attenuation, differential PC and DF channels on biological samples \cite{Brombal2023PEPI}. Similarly to the multimodal X-ray system involved in the work of Olivo et al., the equipment installed at PEPILab features specifications meeting the depiction of AC structures. In the perspective of preclinical applications, a CT prototype has been developed to deliver USAXS images. Its application to lung evaluation on healthy humans and patients with pulmonary disorders yielded results comparable to conventional radiography \cite{Willer2021X,Gassert2021X}. Analogously, applications of DF could be extended to musculoskeletal imaging. For instance, articular soft tissues might be distinguished, as well as their alteration, thanks to the subpixel spatial resolution inferred by gratings configuration.

\section{Conclusions}

The present scoping review focused on X-ray techniques adopted for AC imaging. One of the main differentiations accounts for the prompt implementation of these approaches with either commercial systems or synchrotron-driven setups. In the first case, CE techniques can be easily implemented with radiopaque CAs. Nonetheless, the use of CAs might alter the pristine properties of the tissue. Conversely, PC methods do not imply any alteration of AC nature, but require stringent specifications for the X-ray source. The sought-after coherence of the X-ray beam, crucial for PC imaging, is achieved with SR. Thanks to the latter implementation, various optical configurations retrieve valuable information of AC, down to the cellular scale. Regardless of the intrinsic contrast enhancement associated with PC, the limited access to SR facilities hinders the full unfolding of refraction-based imaging modalities. A promising option makes use of conventional X-ray sources, together with suitable optical elements. Still, the low brilliance of conventional sources significantly increases the acquisition times. Lastly, the scattering information associated with DF is still under investigation for various biomedical applications, including the musculoskeletal system.

The context becomes further intricated in case rheologic studies are considered. Among the approaches here discussed, PC methods provide visual and quantitative information. According to the existing literature, PBPC and GI successfully depict the inner structures of AC and enable their implementation to DVC analysis. More in detail, PBPC features the shortest acquisition time to resolve the microscopic lacunae. On the contrary, GI and, in general, PC methods based on optical elements leads to increased exposure times, raising concerns on the deposited radiation dose and sample stability. At present, this limits rheologic evaluation of AC samples mostly to synchrotron-based experiments. However, the development of novel synchrotron-like X-ray sources with sources with a smaller footprint and cost, such as laser-driven system of liquid-metal jet sources, hold the promise to impact AC imaging, enabling advanced applications, such as rheology, in compact laboratory environment.



\vspace{1cm}

\textbf{Authors' contributions: }{S.F.: Conceptualization; Data curation; Formal analysis; Investigation; Methodology; Project administration; Software; Supervision; Validation; Visualization; Writing – original draft; Writing – review and editing. L.B.: Conceptualization; Data curation; Formal analysis; Supervision; Validation; Visualization; Writing – review and editing. P.C.: Conceptualization; Data curation; Formal analysis; Supervision; Validation; Visualization; Writing – review and editing. F.B.: Conceptualization; Funding acquisition; Investigation; Methodology; Project administration; Resources; Supervision; Validation; Visualization; Writing – review and editing.}

\vspace{1cm}

\textbf{Acknowledgements: }{We would like to thank Luigi Lena for providing the figures in this paper.}

\vspace{1cm}

\textbf{Funding: }{This research was co-funded by the Italian Complementary National Plan PNC-I.1 "Research initiatives for innovative technologies and pathways in the health and welfare sector” D.D. 931 of 06/06/2022, "DARE - DigitAl lifelong pRevEntion" initiative, code PNC0000002, CUP: B53C22006230001.}


\vskip2pc

\bibliographystyle{abbrv} 
\bibliography{ref}

\small

\end{document}